\documentclass[%
 aip,
rsi,%
 amsmath,amssymb,
 reprint,%
]{revtex4-1}

\usepackage{graphicx}
\usepackage{dcolumn}
\usepackage{bm}

\begin{document}


\title{A Coincidence Velocity Map Imaging Spectrometer for Ions and High-Energy Electrons to Study Inner-Shell Photoionization of Gas-Phase Molecules}
 
\author{Utuq Ablikim}
\affiliation{J.R.~Macdonald Laboratory, Department of Physics, Kansas State University, Manhattan, KS 66506, USA}
\affiliation{Advanced Light Source, Lawrence Berkeley National Laboratory, Berkeley, CA 94720, USA}
\author{C\'edric Bomme}
\affiliation{Deutsches Elektronen-Synchrotron (DESY), 22607 Hamburg, Germany}
\author{Timur Osipov}
\affiliation{SLAC National Accelerator Laboratory, Menlo Park, CA 94025, USA}
\author{Hui Xiong}
\affiliation{Department of Physics, University of Connecticut, Storrs, CT 06269, USA}
\author{Razib Obaid}
\affiliation{Department of Physics, University of Connecticut, Storrs, CT 06269, USA}
\author{Ren\'e C. Bilodeau} 
\affiliation{Advanced Light Source, Lawrence Berkeley National Laboratory, Berkeley, CA 94720, USA}
\affiliation{Department of Physics, University of Connecticut, Storrs, CT 06269, USA}
\author{Nora G. Kling}
\affiliation{Department of Physics, University of Connecticut, Storrs, CT 06269, USA}
\author{Ileana Dumitriu}
\affiliation{Hobart and William Smith Colleges, Geneva, NY 14456, USA}
\author{Sven Augustin}
\affiliation{J.R.~Macdonald Laboratory, Department of Physics, Kansas State University, Manhattan, KS 66506, USA}
\affiliation{Max Planck Institute for Nuclear Physics, 69117 Heidelberg, Germany}
\author{Shashank Pathak}
\affiliation{J.R.~Macdonald Laboratory, Department of Physics, Kansas State University, Manhattan, KS 66506, USA}
\author{Kirsten Schnorr}
\affiliation{Max Planck Institute for Nuclear Physics, 69117 Heidelberg, Germany}
\affiliation{Department of Chemistry, University of California, Berkeley, CA 94720, USA}
\author{David Kilcoyne} 
\affiliation{Advanced Light Source, Lawrence Berkeley National Laboratory, Berkeley, CA 94720, USA}
\author{Nora Berrah}
\affiliation{Department of Physics, University of Connecticut, Storrs, CT 06269, USA}
\author{Daniel Rolles}
\email{rolles@phys.ksu.edu}
\affiliation{J.R.~Macdonald Laboratory, Department of Physics, Kansas State University, Manhattan, KS 66506, USA}

\date{\today}

\begin{abstract}
We report on the design and performance of a double-sided coincidence velocity map imaging spectrometer optimized for electron-ion and ion-ion coincidence experiments studying inner-shell photoionization of gas-phase molecules with soft X-ray synchrotron radiation. The apparatus employs two microchannel plate detectors equipped with delay-line anodes for coincident, time- and position-resolved detection of photo- and Auger electrons with kinetic energies up to 300\,eV on one side of the spectrometer and photoions up to 25\,eV per unit charge on the opposite side. We demonstrate its capabilities by measuring valence photoelectron and ion spectra of neon and nitrogen, and by studying channel-resolved photoelectron and Auger spectra along with fragment-ion momentum correlations for chlorine $2p$ inner-shell ionization of \textit{cis}- and \textit{trans}-1,2-dichloroethene. 
\end{abstract}


.\maketitle

\section{\label{sec:intro}INTRODUCTION}

Since the pioneering work of Houston and Chandler~\cite{chandler_two-dimensional_1987}
and Eppink and Parker~\cite{eppink_velocity_1997}, charged-particle imaging and, in particular, velocity map imaging (VMI) have become a wide-spread technique to study photoionization, dissociation, and gas-phase chemical reactions dynamics in a wide variety of targets and with a number of different light sources.~\cite{chandler_perspective:_2017,agsuites2001,KR2012} Common applications of the VMI technique range from experiments with laboratory sources such as gas-discharge lamps, picosecond and femtosecond laser systems, and high harmonic generation sources to studies performed on large-scale facilities such as synchrotron radiation sources and free-electron lasers. Velocity map imaging is particularly important for measurements with extreme ultraviolet (XUV) and X-ray photons since the radiation can ionize along the entire beam path, not just in a high-intensity focal region, as is it the case, e.g., for femtosecond near-infrared lasers. In order to obtain good energy resolution, it is therefore crucial to focus charged particles with the same velocity but originating from different starting points in the interaction region to the same spot on the detector. 

In parallel to the emergence of the VMI, 
coincidence momentum imaging methods, such as COLTRIMS~\cite{dorner_cold_2000,ullrich_recoil-ion_2003,ullrich_recoil-ion_1997} and related techniques~\cite{Davies1999,lafosse2000,DOWEK2002323}, also developed into powerful tools for imaging gas-phase photoionization and photofragmentation dynamics. In recent years, several "hybrid" spectrometers have been developed that use elements of both traditional VMI and COLTRIMS-style spectrometers, thereby overcoming some of the limitations of each of the techniques. Examples of such double-sided hybrid spectrometers for coincident electron and ion detection include combinations of a VMI spectrometer with a traditional time-of-flight or electrostatic cylindrical analyzer opposite to the VMI side~\cite{garcia_two-dimensional_2004,garcia_refocusing_2005, rolles_velocity_2007,pesic_velocity_2007, red_exploring_2010,Bodi2009,POKeeffe2011}, as well as double-sided VMI spectrometers with two position-sensitive detectors for coincident momentum-resolved imaging of both, electrons and ions~\cite{Strueder2010,rolles_femtosecond_2014,Takahashi2000,Vredenborg2008}, many of which were specifically designed for photoionization studies using synchrotron radiation~\cite{garcia_delicious_2013,bomme_double_2013,sztaray_crf-pepico:_2017,Hosaka2006,Xiaofeng2009,bodi2012}. With few exceptions~\cite{Hosaka2006}, the latter are mostly optimized for and limited to rather low-kinetic energy electrons. Here, we present the design of a double-sided VMI spectrometer capable of detecting high-energy electrons up to 300 eV in order to allow for electron-ion and ion-ion coincidence measurements~\cite{Ablikim2016, Ablikim2017, HuiXiong2017} and for the angle-resolved detection of Auger electrons produced by inner-shell photoionization with soft X-ray synchrotron radiation. As described in detail in section~\ref{sec:results}, this enables, e.g., detection of Auger electrons in coincidence with momentum-resolved fragment ions in order to measure channel-resolved Auger spectra, thereby disentangling congested molecular Auger spectra into individual contributions from specific ionic final states.  

Instead of a double-sided design as presented here, an alternative way of realizing electron-ion coincidence experiments with a VMI spectrometer is to use a single-sided spectrometer where the extraction fields are switched after the electron detection in order to detect the ions on the same detector~\cite{Lehmann2012, Zhao2017, Fan2017}. While the latter allows for a fully independent choice of extraction fields for electrons and ions and can reduce the cost of the apparatus since only one position-sensitive detector is needed, the high-voltage switching typically generates a large amount of ringing on the detector that can make it difficult or impossible to detect ions with short flight times, such as H$^+$. In general, it also reduces the momentum resolution of the ions, although the actual amount depends on the quality of the high-voltage pulse and other details of the experiment.

\section{\label{apparatus}APPARATUS}
\subsection{\label{sec:chamber}Vacuum system and molecular beam setup}
The ultrahigh vacuum (UHV) system of the coincidence end-station consists of three differentially pumped chambers: (i) The interaction chamber containing the double-sided VMI spectrometer; (ii) the molecular beam source; and (iii) a differential pumping section to decouple the experimental chamber from the beamline vacuum. The interaction chamber is pumped by two turbomolecular pumps (600\,l/s and 360 l/s) and typically reaches a base pressure in the low 10$^{-8}$\,mbar without bake-out. It has an internal double $\mu$-metal shielding to minimize the influence of the earth's magnetic field and other stray magnetic fields on the electron trajectories. The continuous molecular beam is mounted on an xyz-manipulator and consists of a stainless steel tube terminated by 30 micron flat aperture that can be cooled by liquid nitrogen or (moderately) heated by a resistive heater. The molecular beam expansion chamber is pumped by two 1000\,l/s turbomolecular pumps. Typically, a single skimmer with 500 $\mu$m diameter separates the expansion chamber from the interaction chamber, but a second pumping stage with an additional skimmer can be added, depending on the experimental requirements.

\subsection{\label{sec:spectrometer}Spectrometer design and performance}
Although the VMI apparatus can, in principle, be used at different light sources and different beamlines, it was designed with the parameters of beamline 10.0.1.3 of the Advanced Light Source at the Lawrence Berkeley National Laboratory in mind since it is most frequently being used at this beamline. Beamline 10.0.1.3 delivers VUV and soft X-ray radiation in the photon energy range between 17\,eV and 350\,eV with a resolving power of 35,000 at 60\,eV~\cite{kaindl_ultrahigh_1995, warwick_performance_1995}, covering, for example, the first inner-shell ionization thresholds of many halogens such as chlorine, bromine, and iodine, the L-shell ionization threshold of sulfur, and the K-shell ionization threshold of carbon. In order to be able to detect all photo- and Auger electrons over the entire photon energy range, it is therefore necessary that the spectrometer can detect electrons up to 300\,eV with a full-solid-angle acceptance. Furthermore, since the focus of our research lies on the photoionization and fragmentation dynamics of molecules after inner-shell photoionization, typical fragment ion kinetic energies can reach up to $\approx$ 25\,eV, and up to three or even more ionic fragments need to be detected in coincidence and with good time and position resolution in order to be able to reconstruct their three-dimensional momentum vectors.
\begin{figure}
 \centering
 \includegraphics[width=\linewidth]{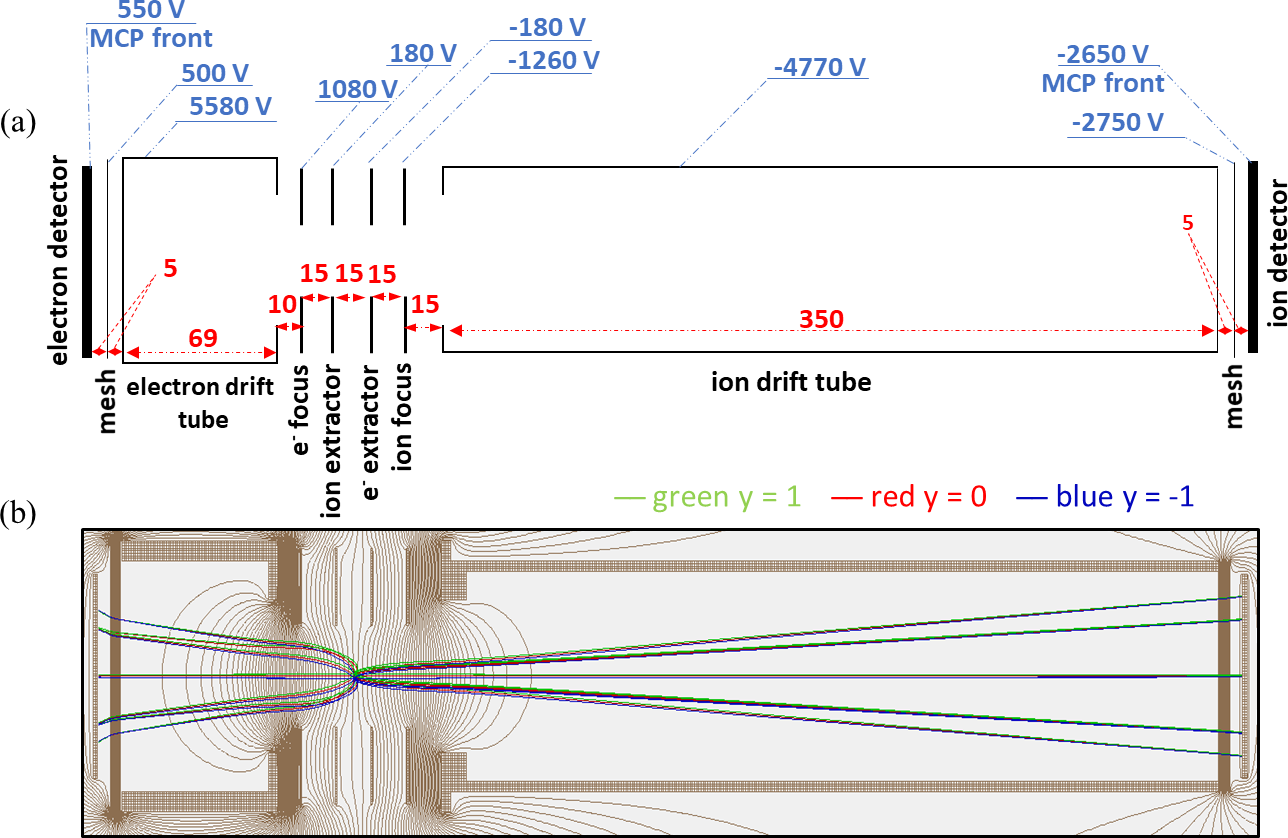}
 \caption{(a) Schematics of the double-sided VMI spectrometer including the geometric dimensions (length unit: mm) and typical voltages for the spectrometer electrodes and the detectors. (b) SIMION equipotential lines and simulated trajectories of ions with a kinetic energy of 15\,eV and electrons with a kinetic energy of 120\,eV emitted into five different directions, towards and away from the detector and at 0, 45, 90, and 135 degrees with respect to the spectrometer axis, and at three positions along the direction of the molecular beam, each spaced 1 mm apart (green, red, and blue lines, respectively).}
 \label{fgr:spec}
\end{figure}

Since the electron and the ion side in a double-sided VMI spectrometer share a common extraction field, their performance is interconnected. Therefore, the lengths of the drift tubes on both sides need to be designed according to the required kinetic energy ranges mentioned above. Fig.~\ref{fgr:spec} shows the dimensions of the spectrometer along with typical voltage settings and the corresponding SIMION simulations. Electrons and ions created in the interaction region located in the middle between the electron and ion extractor electrodes are accelerated by static electric fields towards two opposite MCP detectors with 80 mm diameter. In addition to the two extractor electrodes, two additional electrodes ("electron focus" and "ion focus") provide tunable electrostatic lenses that help focusing the extended interaction region~\cite{rolles_velocity_2007} and help compensate for field distortions due to the open extractor/repeller geometry~\cite{vredenborg_photoelectron-photoion_2008}. Varying the voltage on these focusing electrodes will shift which kinetic energy range is focused best, and thus has the best kinetic energy resolution.
 \begin{figure*}[th]
 \centering
 \includegraphics[width=\linewidth]{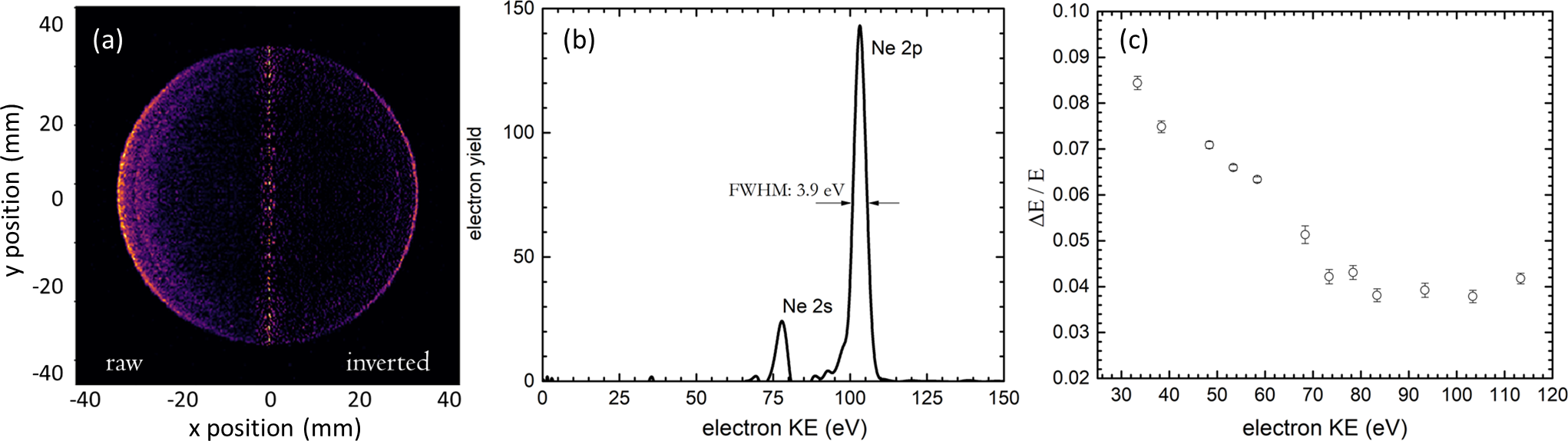}
 \caption{(a) Raw (left) and inverted (right) photoelectron image for photoionization of neon atoms at a photon energy of 125 eV. (b) Corresponding photoelectron spectrum, i.e., 2$\pi$ angle-integrated radial distribution, obtained from the Abel inverted image. (c) Relative kinetic energy (KE) resolution, $\Delta E / E$, as a function of electron KE, determined from the full width at half maximum (FWHM) of the Ne(2$p$) photoelectron peak.}
 \label{fgr:Ne_elec}
\end{figure*}

For the simulation shown here, the electrode voltages of the VMI spectrometer were chosen to allow for the collection of electrons up to 290 eV and ions up to 25 eV per unit charge over the full solid angle. This was achieved by applying +180 V and -180 V to the two inner-most extractor electrodes, +1080 and -1260 V to the two focusing electrodes, and +5580 V/-4770 V to the two drift tubes. 

In order to detect electrons and ions in coincidence and to record both position and time information for the charged fragments, which is necessary to determine their three-dimensional momentum vectors, the double-sided VMI is equipped with microchannel plate (MCP) detectors with multi-hit delay-line anodes (RoentDek DLD80 for the electrons and RoentDek HEX80 for the ions). The analog MCP and delay line signals are amplified, processed by a constant fraction discriminator (CFD), and then recorded by the data acquisition system consisting of two RoentDek TDC8HP 8-channel multi-hit time-to-digital converters (TDC). The TDCs have a resolution of $<$100 ps and a multi-hit dead-time of $<$10 ns. They are triggered by the detection of an electron (which could be either a photoelectron or an Auger electron), which typically arrives at the detector after a flight time of approximately 5 ns. The ion time of flight is then measured with respect to the arrival time of the first detected electron. This allows the experiment to be performed during the standard ALS multi-bunch top-off mode of operation, which has an electron bunch spacing in the storage ring of 2 ns. In the ALS's two-bunch timing mode of operation, the spacing between the soft X-ray pulses is large enough to unambiguously attribute the detected electrons to a specific light pulse, thus making it possible to also determine the electron flight times by using the ALS bunch marker signal as a timing reference.

\subsection{\label{sec:electrons}Two-dimensional electron imaging}
Since the time-of-flight spread of the electrons due to their initial momentum along the spectrometer axis is very small and since the spacing between electron bunches in the storage ring in the multi-bunch operation mode is not sufficient to link the detected photo- or Auger electrons to a specific soft X-ray pulse, we typically do not attempt to retrieve any momentum information from the electron time of flight. Instead, the electron images are treated as two-dimensional projections of the three-dimensional momentum sphere, and an image inversion procedure is employed to retrieve electron kinetic energy spectra and angular distributions as typically done for "conventional" VMI spectrometers. To demonstrate the performance of the electron side of the double-sided spectrometer and to characterize its kinetic energy resolution, photoelectron images for the valence photoionization of neon atoms were recorded over a range of photon energies between 50 and 150 eV.
\begin{figure}[bh]
 \centering
  \includegraphics[width=\linewidth]{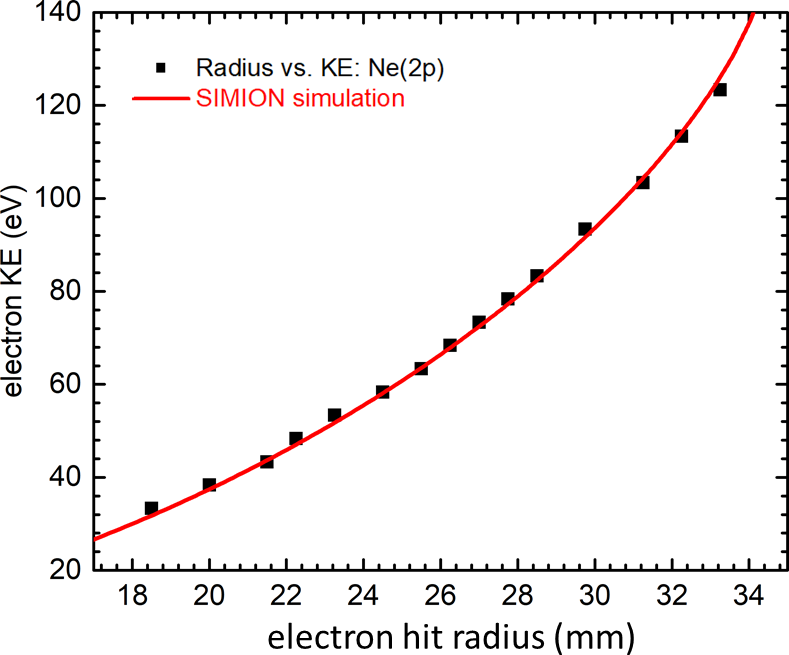}
\caption{Radial position of the Ne(2$p$) (black squares) 
photoelectron lines in the inverted electron images for different electron kinetic energies, compared to the results obtained from SIMION trajectory calculations for the same spectrometer voltages as used in the experiment.}
  \label{fgr:e_calib}
\end{figure}

A raw electron image recorded at a photon energy of 125 eV is shown on the left hand side of Fig.~\ref{fgr:Ne_elec}(a), together with a central slice through the three-dimensional momentum sphere, obtained by applying an inverse Abel transformation~\cite{hickstein_pyabel_nodate}, shown on the right hand side of the same image. The corresponding angle-integrated photoelectron spectrum obtained from the Abel inversion is shown in Fig.~\ref{fgr:Ne_elec}(b). By determining the width of the Ne(2$p$) photoelectron peak as a function of photon energy and thus of photoelectron kinetic energy, the resolution of the electron side of the spectrometer can be characterized, as shown in Fig.~\ref{fgr:Ne_elec}(c). We neglect the line broadening due to the photon bandwidth, which was less than 10 meV for the monochromator exit slit settings chosen for this measurement. Note that in order to correct for distortions of the electron images, which severely affect the achievable kinetic energy resolution, a circularization procedure has been applied to the raw images before the inversion.~\cite{hickstein_pyabel_nodate, Gascooke2017}

Given the known Ne($2p$) and Ne($2s$) ionization potentials of 21.6 eV and 48.5 eV, respectively, the neon measurement can also be used to determine the calibration function for converting the radius of the inverted electron images into electron kinetic energy, as shown in Fig.~\ref{fgr:e_calib}. Experimental data and SIMION trajectory simulations are in good agreement, verifying that the latter can safely be used for calibrating the electron kinetic energies.

\subsection{\label{sec:electrons}Three-dimensional ion momentum imaging}
In order to test and characterize the momentum imaging capabilities of the ion side, we have performed an electron-ion-ion coincidence experiment on the dissociative valence double photoionization of N$_2$ and O$_2$. N$_2$ and O$_2$ molecules were introduced into the interaction chamber as a supersonic jet by directly opening the gas line to ambient air. The spectrometer voltages for this measurement were +100 V and -100 V on the two inner-most extractor electrodes, +600 and -700 V on the two additional focusing lenses, and +3100 V and -2650 V on the two drift tubes.

\begin{figure}[t]
\centering
 \includegraphics[width=\linewidth]{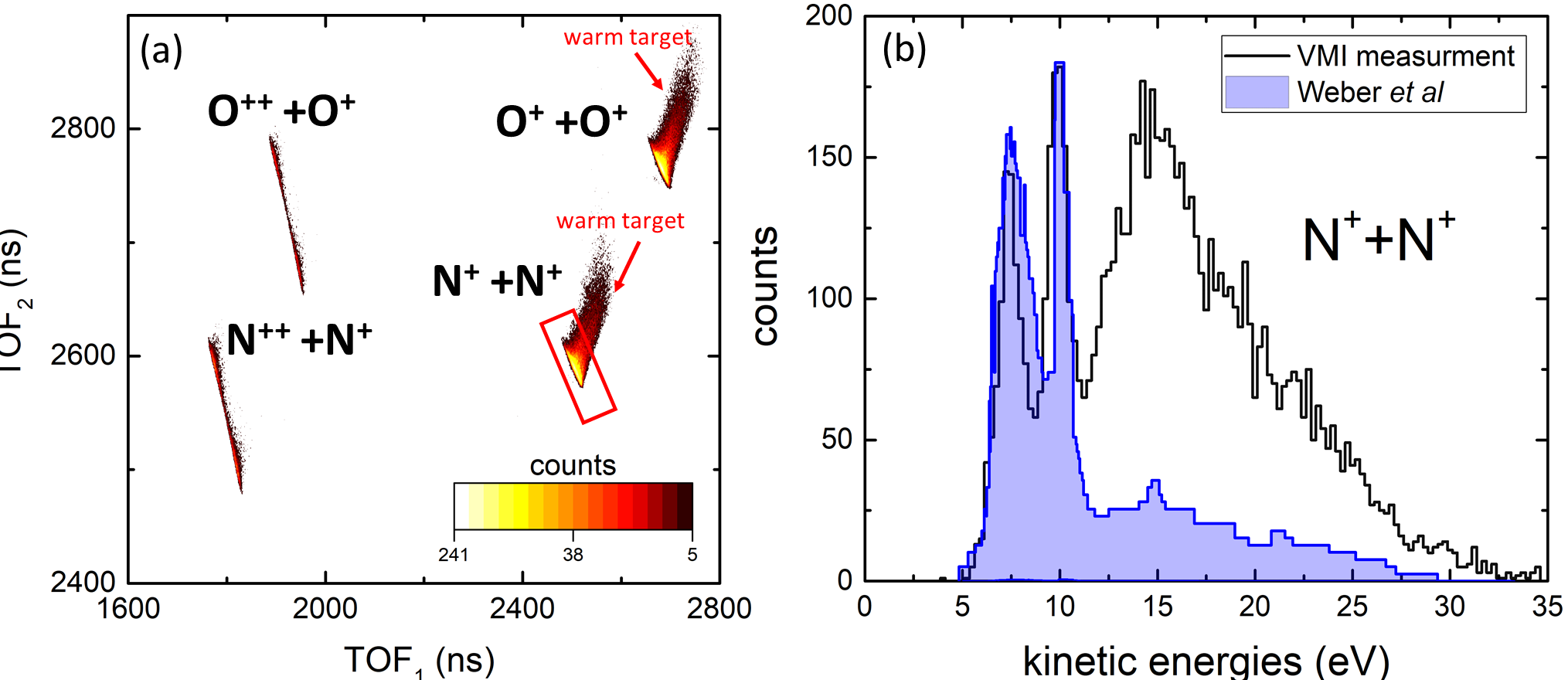}
\caption{ (a) Photoion-photoion coincidence (PIPICO) plot for photoionization of N$_2$ and O$_2$ at 160 eV photon energy. (b) KER distribution of the N$^+$+N$^+$ channel after photoionization at 160 eV photon energy (black line) compared to the KER distribution of the N$^+$+N$^+$ channel produced by N$_2$ inner-shell ionization at 419.3 eV obtained using a COLTRIMS apparatus~\cite{weber_k-shell_2001}.}
  \label{fgr:i_calib}
\end{figure}

Fig.~\ref{fgr:i_calib}(a) shows the resulting photoion-photoion coincidence (PIPICO) plot for photoionization at 160 eV photon energy. The elongated tails towards longer flight times that are visible for each coincidence channel are due to fragmentation of "warm" N$_2$ and O$_2$ molecules that have either leaked through the skimmer from the expansion chamber into the main chamber or that were present as residual gas. They are ionized along the whole light path. The resulting fragment ions are not focused properly and, as confirmed by SIMION simulations, have longer flight times since they are not produced in the center of the spectrometer. However, their contribution can be significantly reduced by selecting a narrow gate in the PIPICO spectrum.  

Using the time-of-flight and the detector hit position, the three-dimensional momentum of each ion can be reconstructed and the total kinetic energy release (KER) spectrum of the N$^+$+N$^+$ channel can be extracted, as shown as a black line in Fig.~\ref{fgr:i_calib}(b). To demonstrate the good momentum resolution that can be achieved by our spectrometer, the KER spectrum is compared to a KER spectrum measured for N$_2$ inner-shell ionization obtained using a COLTRIMS apparatus~\cite{weber_k-shell_2001}. The width of the two sharp peaks at 8 and 10 eV are comparable in both measurements (the FWHM of the peak at 10 eV is 500 meV (FWHM) in the COLTRIMS measurement and 580 meV (FWHM) in the present experiment). Note that the COLTRIMS data was recorded after inner-shell ionization, where the doubly-charged intermediate state is reached via Auger decay, while the present data was recorded after valence ionization, where the doubly-charged intermediate state is reached via a shake-off process, which may lead to different branching ratios. In addition, for the data analysis shown here, only two layers of the HEX detector were used, which leads to some dead time for fragments with identical mass-to-charge ratio. The dead time depends on both, kinetic energies and angular distributions of the fragments and may therefore affect certain parts of the KER spectrum, such as the two sharp peaks, more than other parts.

Since the electric field in a VMI spectrometer is not homogeneous, one cannot derive analytical formulas to reconstruct the ion momenta from the measured time of flight and hit positions of each ion. Instead, we use the \emph{SIMION 8.1} software package to simulate the expected time of flight and hit positions for ions starting in the interaction region with different kinetic energies and emission angles. 

By varying the starting position of the ions in the SIMION simulation within a sphere of 1 mm diameter (which does not reflect the assumed size of the interaction region but rather the uncertainty with which the exact position is known), the uncertainty $\Delta\,E/E$ of the ion kinetic energy calibration is estimated to be on the order of 10\% for ions with a kinetic energy of 1 eV and less than 5\% for 10-eV ions. To verify the energy calibration, the KER spectrum of N$_2$ molecules was measured (see Fig.~\ref{fgr:i_calib}(b)). The measured kinetic energies of several sharp features in this spectrum agree with literature values~\cite{lundqvist_doppler-free_1996,weber_k-shell_2001} to within 4\%, confirming the above estimate of the energy uncertainty. This uncertainty will increase, and the overall performance of the spectrometer will suffer if the interaction region is not well centered between the two extractor electrodes. Careful and precise alignment of the spectrometer with respect to the synchrotron beam is therefore critical. 
\begin{figure*}[t]
 \centering
 \includegraphics[width=\linewidth]{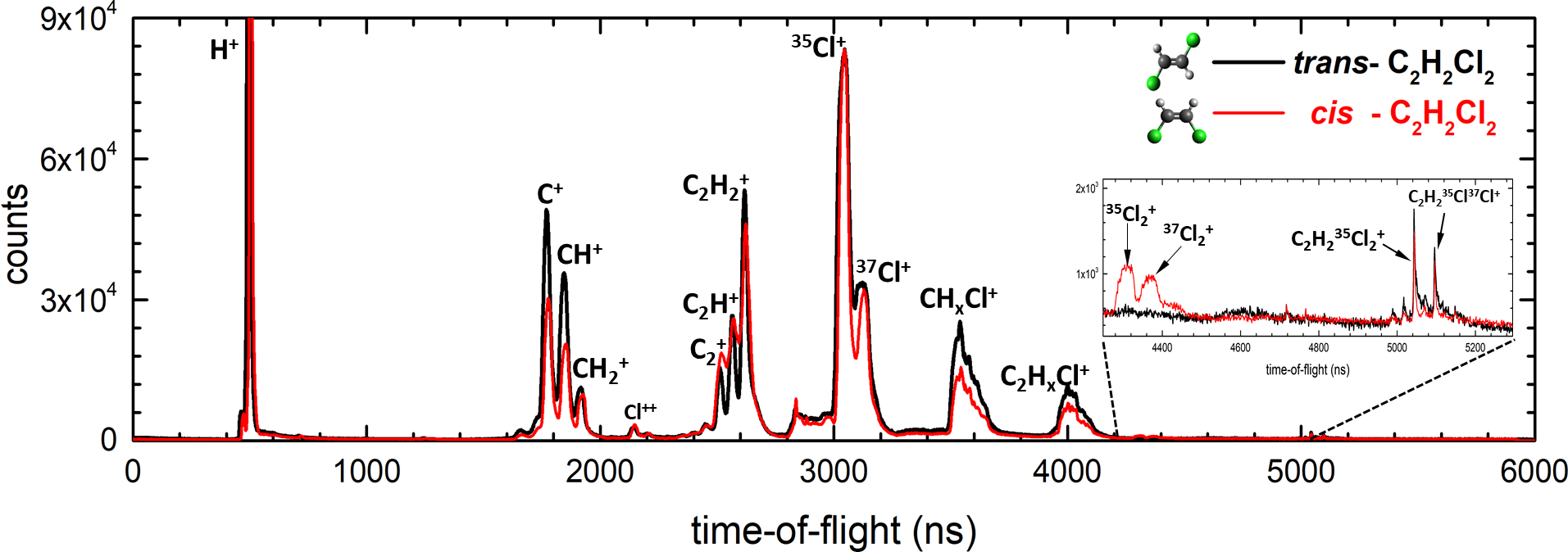}
 \caption{Ion time-of-flight spectrum for inner-shell photoionization of isomerically pure samples of $cis$- (red) and $trans$- (black) 1,2-C$_2$H$_2$Cl$_2$ at a photon energy of 240\,eV. The inset shows a zoom in of the TOF region between 4250 and 5300\,ns. The molecular geometry of $trans$- and $cis$-1,2-C$_2$H$_2$Cl$_2$ is sketched in the top right corner.
 }
 \label{fgr:TOF}
\end{figure*}
\begin{figure*}
 \centering
 \includegraphics[width=\linewidth]{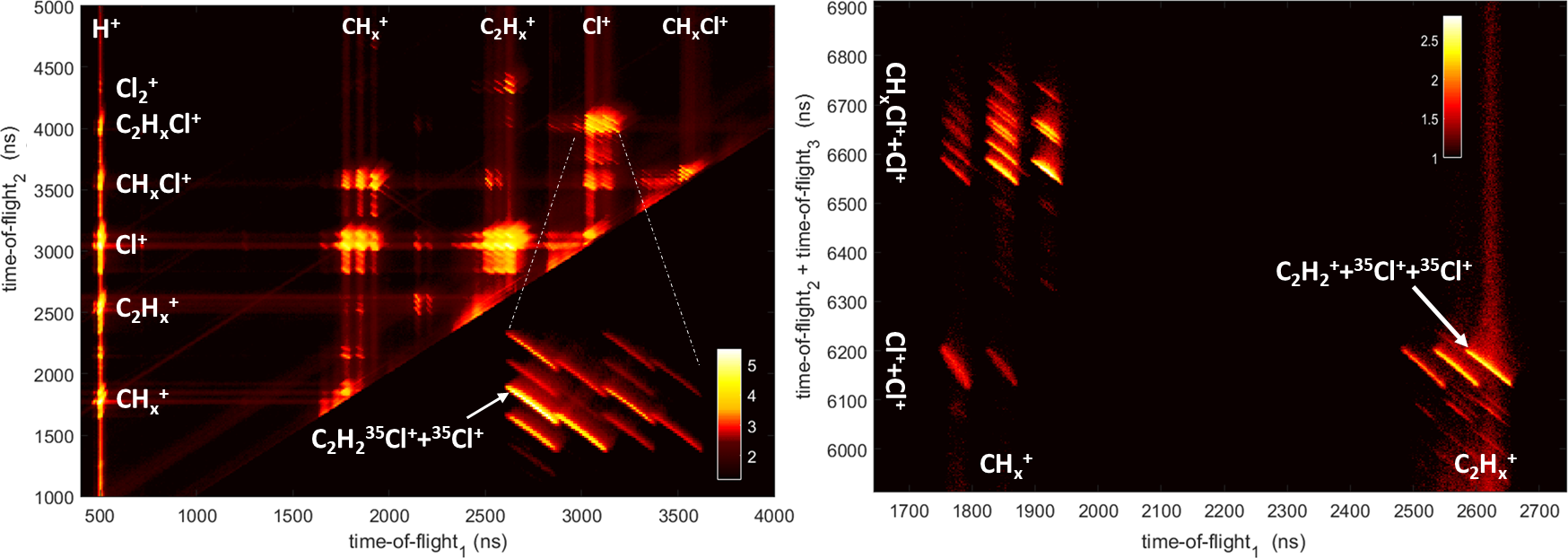}
 \caption{(a) Photoion-photoion coincidence (PIPICO) and (b) photoion-photoion-photoion coincidence (PIPIPICO) maps of $cis$-C$_2$H$_2$Cl$_2$ after inner-shell photoionization at 240\,eV. The coincident ion yields are shown on a logarithmic color scale.}
 \label{fgr:PIPICO}
\end{figure*}


\section{\label{sec:results}Inner-shell ionization of \textit{cis}- and \textit{trans}-1,2-dichloroethene}
In order to demonstrate the capabilities of our double-sided, high-energy coincidence VMI, this section presents results of an electron-ion-ion coincidence experiment studying the inner-shell ionization of $cis$- and $trans$-1,2-C$_2$H$_2$Cl$_2$ molecules in the gas phase. The commercially available, isomerically pure samples (96\% purity, Sigma Aldrich) are liquid at room temperature and were first outgassed in a freeze/pump/thaw cycle before introducing them into the gas phase via expansion through a 30 micron aperture without additional carrier gas. The molecular beam was crossed by a beam of linearly polarized soft X-ray radiation from the ALS (240 eV photon energy with 150 meV bandwidth) in the interaction region of the VMI spectrometer. The lens voltages of the VMI spectrometer were chosen as shown in Fig.~\ref{fgr:spec} to allow for the collection of electrons up to 290 eV, singly charged ions up to 25 eV, and doubly charged ions up to 50 eV over the full solid angle. 

A sketch of the molecular geometry of the $cis$ and $trans$ isomers along with the ion time-of-flight spectra recorded for each isomer at a photon energy of 240\,eV (40\,eV above the Cl($2p$) inner-shell ionization threshold) is shown in Fig.~\ref{fgr:TOF}. The two spectra are normalized to have an equal yield of $^{35}$Cl$^+$ ions and clearly show isomer-dependent differences in the branching ratios, most notably in the yield of the Cl$_2^+$ fragments, which are created by bond recombination and almost exclusively occur after photoionization of $cis$-1,2-C$_2$H$_2$Cl$_2$.

Further information about the fragmentation of 1,2-C$_2$H$_2$Cl$_2$ can be obtained from the PIPICO and the photoion-photoion-photoion coincidence (PIPIPICO) maps shown in Fig.~\ref{fgr:PIPICO}(a) and (b), respectively, for the $cis$ isomer. In the PIPICO map, the time of flight of the second detected ion (time-of-flight$_2$) is plotted against the time of flight of the first detected ion (time-of-flight$_1$). Bright orange and yellow islands mark ionic fragments that arrive in coincidence, i.e., they originate from the same molecule that broke up as a consequence of the ionization event. In particular, sharp diagonal stripes, as shown, e.g., in the zoomed inset in the lower right of Fig.~\ref{fgr:PIPICO}(a), indicate that the molecule broke up into two charged fragments that carry equal momentum, while broader features indicate breakup into more than two fragments with significant momentum. These can be further differentiated in the triple coincidence PIPIPICO map, where the time-of-flight sum of the second and third detected ion (time-of-flight$_2$ + time-of-flight$_3$) is plotted as a function of the time of flight of the first detected ion (time-of-flight$_1$). Note that chlorine has two naturally occurring isotopes, $^{35}$Cl and $^{37}$Cl, which lead to several coincidence channels consisting of all possible combinations of these two isotopes. Additional channels result from the possible loss of one or both hydrogen atoms, all of which can be clearly resolved in the PIPICO and PIPIPICO maps. The strongest complete two-body fragmentation channel, C$_2$H$_2$$^{35}$Cl$^+$+$^{35}$Cl$^+$, and strongest complete three-body fragmentation channel, C$_2$H$_2^+$+$^{35}$Cl$^+$+$^{35}$Cl$^+$, are marked by white arrows in Fig.~\ref{fgr:PIPICO}.

\begin{figure}
 \centering
  \includegraphics[width=\linewidth]{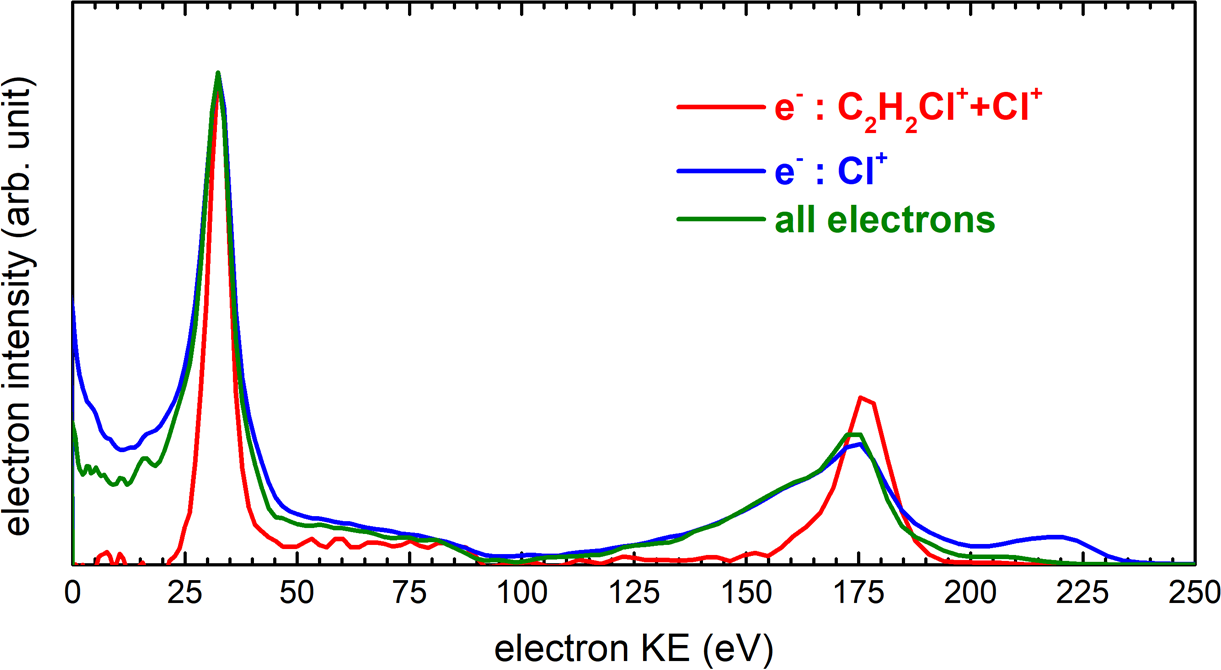}
\caption{Electron kinetic energy spectra recorded at a photon energy of 240 eV in coincidence with the C$_2$H$_2$Cl$^+$+Cl$^+$ channel (red) and in coincidence with a Cl$^+$ ion (blue), compared to the total electron kinetic energy spectrum recorded in coincidence with any possible ion resulting from $cis$-C$_2$H$_2$Cl$_2$ (green). The Cl(2$p$) photoline at $\approx$ 33\,eV is the most prominent spectral feature, while different Auger channels contribute to the broader feature between 125 and 200\,eV. The smaller feature between 200 and 230 eV corresponds to photoelectrons from valence photoionization. All spectra are normalized to the maximum of the photoline.}
  \label{fgr:electrons}
\end{figure}

By selecting only those events where the ions appear in a specific ionic coincidence channel in the PIPICO or PIPIPICO map, it is possible to distinguish photo- and Auger electrons corresponding to different ionic final states and to produce \emph{channel-resolved} photo- and Auger electron spectra, as shown in Fig.~\ref{fgr:electrons}. The width of the Auger feature in the channel-resolved spectrum (red) is considerably narrower than in the non-coincident electron spectrum (green) and the spectrum where only one of the ionic fragments is determined (blue) since the latter two contain a multitude of Auger transitions, while only one or a few specific transitions lead to a given ionic final state. Measurement of electron-ion-ion and electron-ion-ion-ion coincidences can thus help with disentangling convoluted molecular Auger spectra, as will be discussed in more detail in a forthcoming publication.

\begin{figure}
 \centering
  \includegraphics[width=\linewidth]{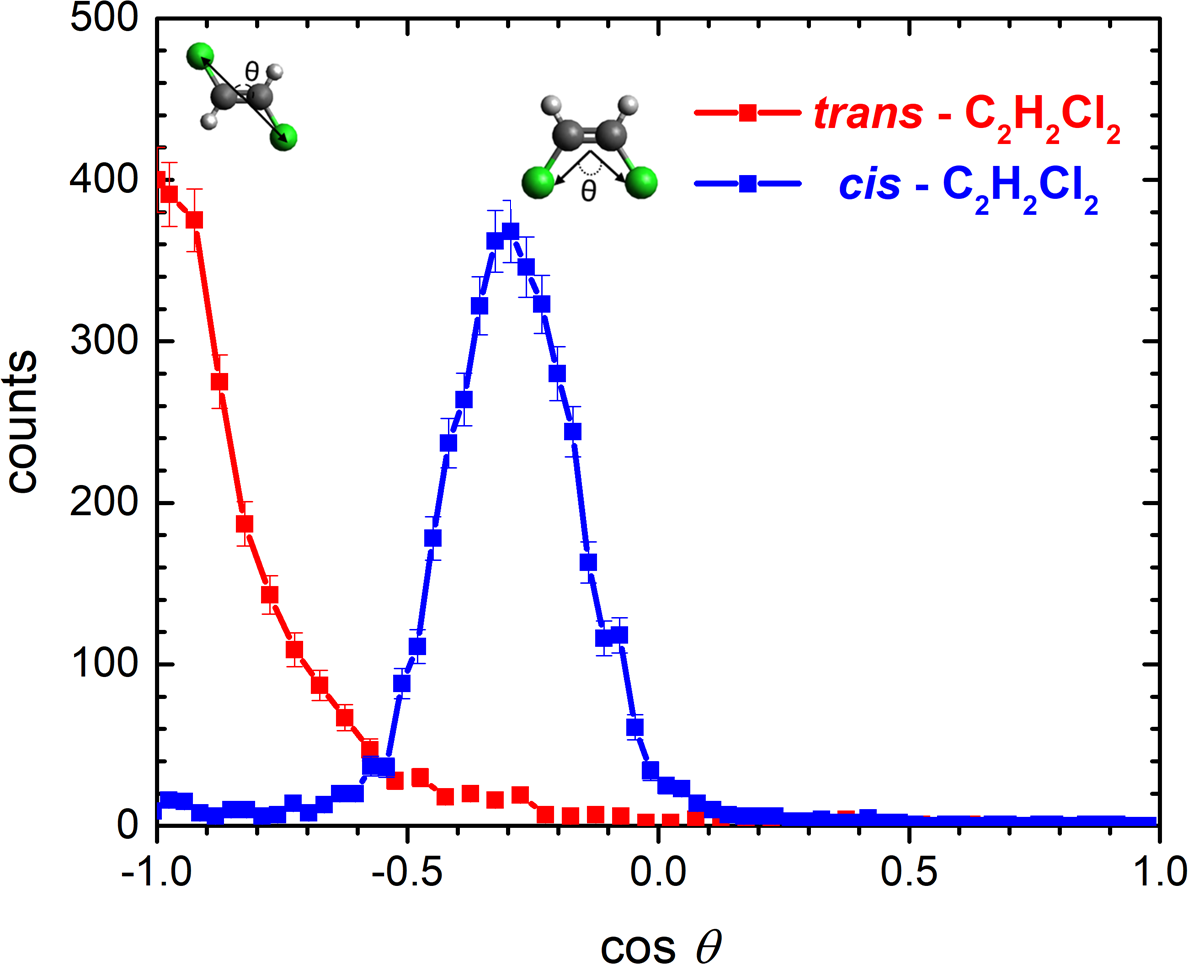}
\caption{Angle between the Cl$^+$ ion momenta in the triple-coincidence channel C$_2$H$_2^+$+Cl$^+$+Cl$^+$ for $cis$- (blue) and $trans$- (red) C$_2$H$_2$Cl$_2$. 
}
  \label{fgr:cos}
\end{figure}

Finally, as we have shown previously~\cite{Ablikim2016}, it is possible to distinguish between different geometric isomers by means of momentum-resolved ion-ion coincidence momentum imaging. For this purpose, we determine the angle $\theta$ between the momentum vectors of two Cl$^+$ fragments detected in coincidence with a C$_2$H$_2$$^+$ fragment. The corresponding yield as a function of the cosine of the angle $\theta$ for both $cis$ and $trans$ isomers is shown in Fig.~\ref{fgr:cos}. For the $trans$ isomers, the distribution peaks at $cos\,\theta = -1$, while it peaks at $cos\,\theta \approx -0.3$ for the $cis$ isomers. Note that the latter value is quite different from our finding in $cis$-C$_2$H$_2$Br$_2$~\cite{Ablikim2016}, where the yield for the $cis$ isomers peaked at $cos\,\theta \approx -0.5$.
This shift to a smaller angle is also reproduced by numerical Coulomb explosion simulation, which assume purely Coulombic repulsion between point charges located at the center of mass of each fragment and instantaneous creation of the charges followed by explosion from the equilibrium geometry of the neutral molecule, as explained in Refs.~\cite{Ablikim2016, Ablikim2017}. The simulations yield value of $cos\,\theta=-0.49$ for the $cis$ isomer of C$_2$H$_2$Cl$_2$ as opposed to a value of $cos\,\theta=-0.58$ for $cis$-C$_2$H$_2$Br$_2$. Note that the simulation overestimates the angle, similar to our findings in the case of C$_2$H$_2$Br$_2$~\cite{Ablikim2016}.

Nevertheless, with the help of these simulations, we can determine that a significant contribution to the difference in the momentum correlations for the $cis$ isomers of C$_2$H$_2$Cl$_2$ and C$_2$H$_2$Br$_2$ is the mass difference between Cl and Br, which changes the energy and momentum sharing with the C$_2$H$_2$$^+$ fragment, while the slightly different equilibrium geometry of C$_2$H$_2$Cl$_2$ and C$_2$H$_2$Br$_2$ affects the simulated angle between the fragments only very little.

\begin{figure}
 \centering
  \includegraphics[width=\linewidth]{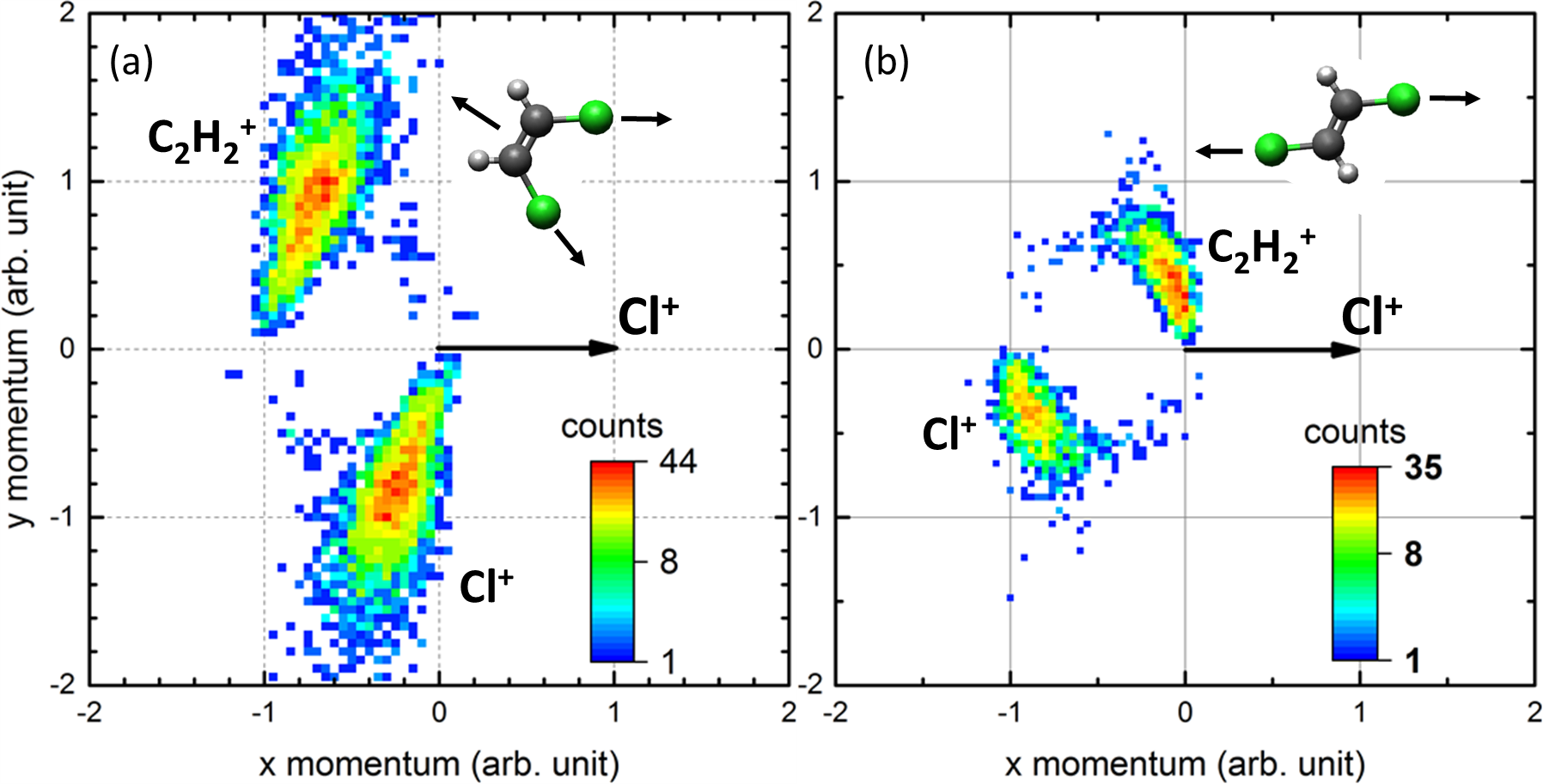}
\caption{Newton plot of the C$_2$H$_2^{+}$+Cl$^{+}$+Cl$^{+}$ triple-coincidence channel for (a) $cis$- and (b) $trans$-C$_2$H$_2$Cl$_2$. The momenta of the C$_2$H$_2^{+}$ fragment (upper half) and of one of the Cl$^{+}$ fragments (lower half) are shown in the frame of the momentum of the second Cl$^{+}$ fragment, which is shown as a black horizontal arrow. The momentum vectors of the C$_2$H$_2^{+}$ fragment and the first Cl$^{+}$ fragment are normalized to the length of momentum vector of the second Cl$^{+}$.}
  \label{fgr:newton}
\end{figure}
Another common way of investigating both the kinematics and the dynamics of a three-body fragmentation process, is by representing the fragment ion momentum correlations in a Newton plot, as shown in Fig.~\ref{fgr:newton}. Here, the momenta of two of the fragments (C$_2$H$_2^{+}$ and Cl$^{+}$) are plotted with respect to the momentum of the third fragment (Cl$^{+}$). For the $trans$ isomer, the Newton plot in Fig.~\ref{fgr:newton}(b) shows that the two Cl$^{+}$ fragments are emitted close to back to back, while the C$_2$H$_2^{+}$ fragment remains almost at rest, as one would expect from an instantaneous fragmentation in the undistorted equilibrium geometry, where the center of mass of the C$_2$H$_2$ moiety is exactly in the middle between the two Cl atoms. The $cis$ isomer, shown in Fig.~\ref{fgr:newton}(a), has a distinctly different fragmentation pattern with the Cl$^{+}$ fragments being emitted at a much steeper angle with respect to each other and the C$_2$H$_2^{+}$ fragment also carrying a significant momentum. In both cases, the shape of the features in the Newton plot suggests that the two bonds that are broken in this particular fragmentation process are broken in a concerted manner, since a sequential breakup with a time interval close to (or larger than) the rotational period of the intermediate would lead to a pronounced circular shape of the features in the Newton plot~\cite{Ablikim2017}, which is clearly absent in the present case.

\begin{figure}
 \centering
  \includegraphics[width=\linewidth]{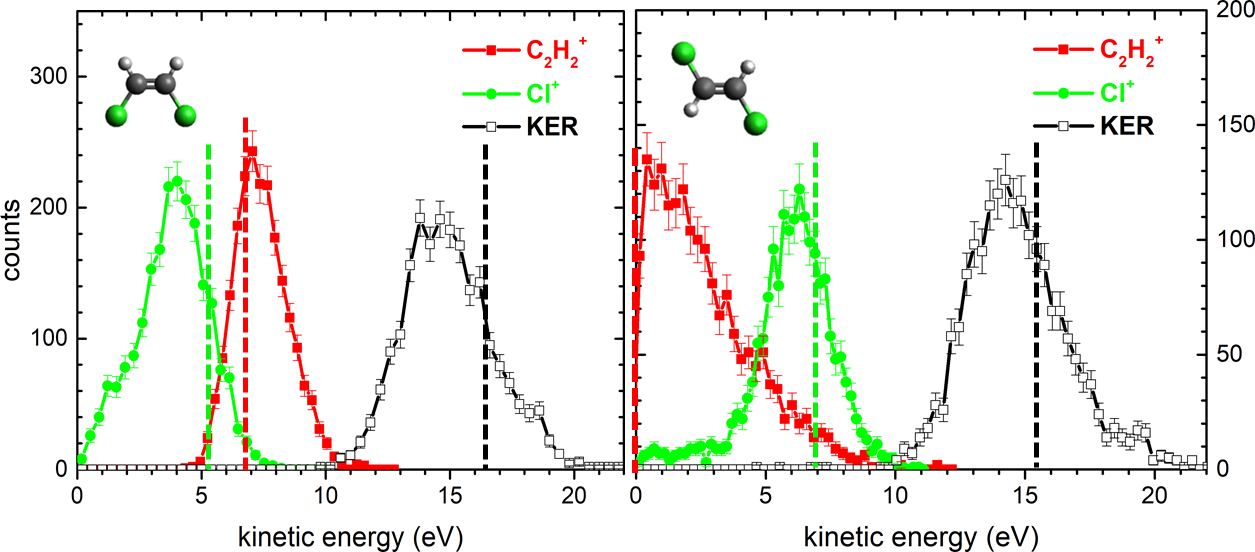}
\caption{Kinetic energy distributions of the Cl$^+$- (green) and C$_2$H$_2^+$-fragments (red) as well as KER (black) in the C$_2$H$_2^+$+Cl$^+$+Cl$^+$ triple-coincidence channel for (a) $cis$ (b) $trans$ isomers. The results of the Coulomb explosion model are shown as dashed lines.}
  \label{fgr:KER}
\end{figure}

The difference in the Coulomb explosion kinematics for $cis$- and $trans$-C$_2$H$_2$Cl$_2$ is also reflected in the kinetic energy distributions and the KER, which are shown in Fig.~\ref{fgr:KER} for the same C$_2$H$_2^+$+Cl$^+$+Cl$^+$ coincidence channel as discussed above. For the $cis$ isomer, the C$_2$H$_2^+$ fragment has significantly higher kinetic energy than each of the Cl$^+$ fragments, while for the $trans$ isomer, the kinetic energy distribution of the C$_2$H$_2^+$ fragment has a maximum at 0 eV and the Cl$^+$ fragments are more energetic than for the $cis$ isomer. Similar to the case of C$_2$H$_2$Br$_2$~\cite{Ablikim2016}, this can be explained by an “obstructed instantaneous explosion”~\cite{Eland1987,ShizukaHsieh1997} in the $trans$ isomer, where the C$_2$H$_2^+$-fragment is trapped between the two Cl$^+$ fragments. This simple picture is corroborated by the results of our Coulomb explosion model calculations shown as dashed lines in Fig.~\ref{fgr:KER}, which reproduce the difference between the fragment kinetic energy distributions in the two isomers. 

\section{\label{sec:Conclusions}CONCLUSIONS}
We have presented the design and performance of a double-sided VMI spectrometer optimized for the coincident detection of high-energy photo- or Auger electrons with energetic ions resulting from the Coulomb explosion of inner-shell ionized molecules. With typical lens voltages up to +/- 6 keV, the spectrometer is capable of time- and position-resolved detection of photo- and Auger electrons with kinetic energies up to 300\,eV and of photoions with up to 25\,eV per unit charge. The kinetic energy acceptance can be increased if higher lens voltages are used. For the typical experimental conditions investigated here, we find a kinetic energy resolution of $\Delta E/E \approx 4 \%$ for electrons with 100 eV kinetic energy and $\approx 5 \%$ for ions with 10 eV kinetic energy. We have illustrated the coincidence capabilities and, in particular, the possibility to record ion coincidence-channel-resolved Auger electron spectra on the example of an Auger electron-ion-ion-ion coincidence experiment performed on \textit{cis}- and \textit{trans}-1,2-dichloroethene at the Advanced Light Source synchrotron radiation storage ring facility.

By combining some of the advantages of traditional VMIs with those of COLTRIMS-type coincidence momentum imaging spectrometers, our double-sided coincidence VMI design is also well suited for high-repetition-rate XUV high harmonic generation sources and for upcoming high-repetition-rate free-electron lasers, where coincident detection of high-energy electron with fragment ions will be an important prerequisit for many gas-phase experiments in atomic and molecular physics and ultrafast photochemistry~\cite{AMO_roadmap}.




\begin{acknowledgments}
This work is supported by the Chemical Sciences, Geosciences, and Biosciences Division, Office of Basic Energy Sciences, Office of Science, U.S.~Department of Energy, Grant No.~DE-FG02-86ER13491 (Kansas State group) and DE-SC0012376 (UConn group). The research used resources of the Advanced Light Source, which is a DOE Office of Science User Facility under contract no.~DE-AC02-05CH11231. We thank the staff of the Advanced Light Source for their hospitality and their support during the beamtimes.
\end{acknowledgments}

\bibliography{aipbib2}

\begin{thebibliography}{44}%
\makeatletter
\providecommand \@ifxundefined [1]{%
 \@ifx{#1\undefined}
}%
\providecommand \@ifnum [1]{%
 \ifnum #1\expandafter \@firstoftwo
 \else \expandafter \@secondoftwo
 \fi
}%
\providecommand \@ifx [1]{%
 \ifx #1\expandafter \@firstoftwo
 \else \expandafter \@secondoftwo
 \fi
}%
\providecommand \natexlab [1]{#1}%
\providecommand \enquote  [1]{``#1''}%
\providecommand \bibnamefont  [1]{#1}%
\providecommand \bibfnamefont [1]{#1}%
\providecommand \citenamefont [1]{#1}%
\providecommand \href@noop [0]{\@secondoftwo}%
\providecommand \href [0]{\begingroup \@sanitize@url \@href}%
\providecommand \@href[1]{\@@startlink{#1}\@@href}%
\providecommand \@@href[1]{\endgroup#1\@@endlink}%
\providecommand \@sanitize@url [0]{\catcode `\\12\catcode `\$12\catcode
  `\&12\catcode `\#12\catcode `\^12\catcode `\_12\catcode `\%12\relax}%
\providecommand \@@startlink[1]{}%
\providecommand \@@endlink[0]{}%
\providecommand \url  [0]{\begingroup\@sanitize@url \@url }%
\providecommand \@url [1]{\endgroup\@href {#1}{\urlprefix }}%
\providecommand \urlprefix  [0]{URL }%
\providecommand \Eprint [0]{\href }%
\providecommand \doibase [0]{http://dx.doi.org/}%
\providecommand \selectlanguage [0]{\@gobble}%
\providecommand \bibinfo  [0]{\@secondoftwo}%
\providecommand \bibfield  [0]{\@secondoftwo}%
\providecommand \translation [1]{[#1]}%
\providecommand \BibitemOpen [0]{}%
\providecommand \bibitemStop [0]{}%
\providecommand \bibitemNoStop [0]{.\EOS\space}%
\providecommand \EOS [0]{\spacefactor3000\relax}%
\providecommand \BibitemShut  [1]{\csname bibitem#1\endcsname}%
\let\auto@bib@innerbib\@empty
\bibitem [{\citenamefont {Chandler}\ and\ \citenamefont
  {Houston}(1987)}]{chandler_two-dimensional_1987}%
  \BibitemOpen
  \bibfield  {author} {\bibinfo {author} {\bibfnamefont {D.~W.}\ \bibnamefont
  {Chandler}}\ and\ \bibinfo {author} {\bibfnamefont {P.~L.}\ \bibnamefont
  {Houston}},\ }\href {\doibase 10.1063/1.453276} {\bibfield  {journal}
  {\bibinfo  {journal} {The Journal of Chemical Physics}\ }\textbf {\bibinfo
  {volume} {87}},\ \bibinfo {pages} {1445} (\bibinfo {year}
  {1987})}\BibitemShut {NoStop}%
\bibitem [{\citenamefont {Eppink}\ and\ \citenamefont
  {Parker}(1997)}]{eppink_velocity_1997}%
  \BibitemOpen
  \bibfield  {author} {\bibinfo {author} {\bibfnamefont {A.~T. J.~B.}\
  \bibnamefont {Eppink}}\ and\ \bibinfo {author} {\bibfnamefont {D.~H.}\
  \bibnamefont {Parker}},\ }\href {\doibase 10.1063/1.1148310} {\bibfield
  {journal} {\bibinfo  {journal} {Review of Scientific Instruments}\ }\textbf
  {\bibinfo {volume} {68}},\ \bibinfo {pages} {3477} (\bibinfo {year}
  {1997})}\BibitemShut {NoStop}%
\bibitem [{\citenamefont {Chandler}, \citenamefont {Houston},\ and\
  \citenamefont {Parker}(2017)}]{chandler_perspective:_2017}%
  \BibitemOpen
  \bibfield  {author} {\bibinfo {author} {\bibfnamefont {D.~W.}\ \bibnamefont
  {Chandler}}, \bibinfo {author} {\bibfnamefont {P.~L.}\ \bibnamefont
  {Houston}}, \ and\ \bibinfo {author} {\bibfnamefont {D.~H.}\ \bibnamefont
  {Parker}},\ }\href {\doibase 10.1063/1.4983623} {\bibfield  {journal}
  {\bibinfo  {journal} {The Journal of Chemical Physics}\ }\textbf {\bibinfo
  {volume} {147}},\ \bibinfo {pages} {013601} (\bibinfo {year}
  {2017})}\BibitemShut {NoStop}%
\bibitem [{\citenamefont {Suits}\ and\ \citenamefont
  {Continetti}(2001)}]{agsuites2001}%
  \BibitemOpen
  \bibfield  {author} {\bibinfo {author} {\bibfnamefont {A.~G.}\ \bibnamefont
  {Suits}}\ and\ \bibinfo {author} {\bibfnamefont {R.~E.}\ \bibnamefont
  {Continetti}},\ }\href@noop {} {\emph {\bibinfo {title} {Imaging in Chemical
  Dynamics}}},\ ACS Symposium Volume 770\ (\bibinfo  {publisher} {American
  Chemical Society},\ \bibinfo {year} {2001})\BibitemShut {NoStop}%
\bibitem [{\citenamefont {Reid}(2012)}]{KR2012}%
  \BibitemOpen
  \bibfield  {author} {\bibinfo {author} {\bibfnamefont {K.~L.}\ \bibnamefont
  {Reid}},\ }\href@noop {} {\bibfield  {journal} {\bibinfo  {journal}
  {Molecular Physics}\ }\textbf {\bibinfo {volume} {110}},\ \bibinfo {pages}
  {131} (\bibinfo {year} {2012})}\BibitemShut {NoStop}%
\bibitem [{\citenamefont {D{\"o}rner}\ \emph {et~al.}(2000)\citenamefont
  {D{\"o}rner}, \citenamefont {Mergel}, \citenamefont {Jagutzki}, \citenamefont
  {Spielberger}, \citenamefont {Ullrich}, \citenamefont {Moshammer},\ and\
  \citenamefont {Schmidt-B{\"o}cking}}]{dorner_cold_2000}%
  \BibitemOpen
  \bibfield  {author} {\bibinfo {author} {\bibfnamefont {R.}~\bibnamefont
  {D{\"o}rner}}, \bibinfo {author} {\bibfnamefont {V.}~\bibnamefont {Mergel}},
  \bibinfo {author} {\bibfnamefont {O.}~\bibnamefont {Jagutzki}}, \bibinfo
  {author} {\bibfnamefont {L.}~\bibnamefont {Spielberger}}, \bibinfo {author}
  {\bibfnamefont {J.}~\bibnamefont {Ullrich}}, \bibinfo {author} {\bibfnamefont
  {R.}~\bibnamefont {Moshammer}}, \ and\ \bibinfo {author} {\bibfnamefont
  {H.}~\bibnamefont {Schmidt-B{\"o}cking}},\ }\href@noop {} {\bibfield
  {journal} {\bibinfo  {journal} {Physics Reports}\ }\textbf {\bibinfo {volume}
  {330}},\ \bibinfo {pages} {95} (\bibinfo {year} {2000})}\BibitemShut
  {NoStop}%
\bibitem [{\citenamefont {Ullrich}\ \emph {et~al.}(2003)\citenamefont
  {Ullrich}, \citenamefont {Moshammer}, \citenamefont {Dorn}, \citenamefont
  {D{\"o}rner}, \citenamefont {Schmidt},\ and\ \citenamefont
  {Schmidt-B{\"o}cking}}]{ullrich_recoil-ion_2003}%
  \BibitemOpen
  \bibfield  {author} {\bibinfo {author} {\bibfnamefont {J.}~\bibnamefont
  {Ullrich}}, \bibinfo {author} {\bibfnamefont {R.}~\bibnamefont {Moshammer}},
  \bibinfo {author} {\bibfnamefont {A.}~\bibnamefont {Dorn}}, \bibinfo {author}
  {\bibfnamefont {R.}~\bibnamefont {D{\"o}rner}}, \bibinfo {author}
  {\bibfnamefont {L.~P.~H.}\ \bibnamefont {Schmidt}}, \ and\ \bibinfo {author}
  {\bibfnamefont {H.}~\bibnamefont {Schmidt-B{\"o}cking}},\ }\href
  {http://iopscience.iop.org/0034-4885/66/9/203} {\bibfield  {journal}
  {\bibinfo  {journal} {Reports on Progress in Physics}\ }\textbf {\bibinfo
  {volume} {66}},\ \bibinfo {pages} {1463} (\bibinfo {year}
  {2003})}\BibitemShut {NoStop}%
\bibitem [{\citenamefont {Ullrich}\ \emph {et~al.}(1997)\citenamefont
  {Ullrich}, \citenamefont {Moshammer}, \citenamefont {D{\"o}rner},
  \citenamefont {Jagutzki}, \citenamefont {Mergel}, \citenamefont
  {Schmidt-B{\"o}cking},\ and\ \citenamefont
  {Spielberger}}]{ullrich_recoil-ion_1997}%
  \BibitemOpen
  \bibfield  {author} {\bibinfo {author} {\bibfnamefont {J.}~\bibnamefont
  {Ullrich}}, \bibinfo {author} {\bibfnamefont {R.}~\bibnamefont {Moshammer}},
  \bibinfo {author} {\bibfnamefont {R.}~\bibnamefont {D{\"o}rner}}, \bibinfo
  {author} {\bibfnamefont {O.}~\bibnamefont {Jagutzki}}, \bibinfo {author}
  {\bibfnamefont {V.}~\bibnamefont {Mergel}}, \bibinfo {author} {\bibfnamefont
  {H.}~\bibnamefont {Schmidt-B{\"o}cking}}, \ and\ \bibinfo {author}
  {\bibfnamefont {L.}~\bibnamefont {Spielberger}},\ }\href {\doibase
  10.1088/0953-4075/30/13/006} {\bibfield  {journal} {\bibinfo  {journal}
  {Journal of Physics B: Atomic, Molecular and Optical Physics}\ }\textbf
  {\bibinfo {volume} {30}},\ \bibinfo {pages} {2917} (\bibinfo {year}
  {1997})}\BibitemShut {NoStop}%
\bibitem [{\citenamefont {Davies}\ \emph {et~al.}(1999)\citenamefont {Davies},
  \citenamefont {LeClaire}, \citenamefont {Continetti},\ and\ \citenamefont
  {Hayden}}]{Davies1999}%
  \BibitemOpen
  \bibfield  {author} {\bibinfo {author} {\bibfnamefont {J.~A.}\ \bibnamefont
  {Davies}}, \bibinfo {author} {\bibfnamefont {J.~E.}\ \bibnamefont
  {LeClaire}}, \bibinfo {author} {\bibfnamefont {R.~E.}\ \bibnamefont
  {Continetti}}, \ and\ \bibinfo {author} {\bibfnamefont {C.~C.}\ \bibnamefont
  {Hayden}},\ }\href {\doibase 10.1063/1.479248} {\bibfield  {journal}
  {\bibinfo  {journal} {The Journal of Chemical Physics}\ }\textbf {\bibinfo
  {volume} {111}},\ \bibinfo {pages} {1} (\bibinfo {year} {1999})}\BibitemShut
  {NoStop}%
\bibitem [{\citenamefont {Lafosse}\ \emph {et~al.}(2000)\citenamefont
  {Lafosse}, \citenamefont {Lebech}, \citenamefont {Brenot}, \citenamefont
  {Guyon}, \citenamefont {Jagutzki}, \citenamefont {Spielberger}, \citenamefont
  {Vervloet}, \citenamefont {Houver},\ and\ \citenamefont
  {Dowek}}]{lafosse2000}%
  \BibitemOpen
  \bibfield  {author} {\bibinfo {author} {\bibfnamefont {A.}~\bibnamefont
  {Lafosse}}, \bibinfo {author} {\bibfnamefont {M.}~\bibnamefont {Lebech}},
  \bibinfo {author} {\bibfnamefont {J.~C.}\ \bibnamefont {Brenot}}, \bibinfo
  {author} {\bibfnamefont {P.~M.}\ \bibnamefont {Guyon}}, \bibinfo {author}
  {\bibfnamefont {O.}~\bibnamefont {Jagutzki}}, \bibinfo {author}
  {\bibfnamefont {L.}~\bibnamefont {Spielberger}}, \bibinfo {author}
  {\bibfnamefont {M.}~\bibnamefont {Vervloet}}, \bibinfo {author}
  {\bibfnamefont {J.~C.}\ \bibnamefont {Houver}}, \ and\ \bibinfo {author}
  {\bibfnamefont {D.}~\bibnamefont {Dowek}},\ }\href {\doibase
  10.1103/PhysRevLett.84.5987} {\bibfield  {journal} {\bibinfo  {journal}
  {Physical Review Letters}\ }\textbf {\bibinfo {volume} {84}},\ \bibinfo
  {pages} {5987} (\bibinfo {year} {2000})}\BibitemShut {NoStop}%
\bibitem [{\citenamefont {Dowek}\ \emph {et~al.}(2002)\citenamefont {Dowek},
  \citenamefont {Brenot}, \citenamefont {Guyon}, \citenamefont {Houver},
  \citenamefont {Lafosse}, \citenamefont {Lebech}, \citenamefont {Jagutzki},\
  and\ \citenamefont {Spielberger}}]{DOWEK2002323}%
  \BibitemOpen
  \bibfield  {author} {\bibinfo {author} {\bibfnamefont {D.}~\bibnamefont
  {Dowek}}, \bibinfo {author} {\bibfnamefont {J.}~\bibnamefont {Brenot}},
  \bibinfo {author} {\bibfnamefont {P.}~\bibnamefont {Guyon}}, \bibinfo
  {author} {\bibfnamefont {J.}~\bibnamefont {Houver}}, \bibinfo {author}
  {\bibfnamefont {A.}~\bibnamefont {Lafosse}}, \bibinfo {author} {\bibfnamefont
  {M.}~\bibnamefont {Lebech}}, \bibinfo {author} {\bibfnamefont
  {O.}~\bibnamefont {Jagutzki}}, \ and\ \bibinfo {author} {\bibfnamefont
  {L.}~\bibnamefont {Spielberger}},\ }\href@noop {} {\bibfield  {journal}
  {\bibinfo  {journal} {Nuclear Instruments and Methods in Physics Research
  Section A: Accelerators, Spectrometers, Detectors and Associated Equipment}\
  }\textbf {\bibinfo {volume} {477}},\ \bibinfo {pages} {323 } (\bibinfo {year}
  {2002})}\BibitemShut {NoStop}%
\bibitem [{\citenamefont {Garcia}, \citenamefont {Nahon},\ and\ \citenamefont
  {Powis}(2004)}]{garcia_two-dimensional_2004}%
  \BibitemOpen
  \bibfield  {author} {\bibinfo {author} {\bibfnamefont {G.~A.}\ \bibnamefont
  {Garcia}}, \bibinfo {author} {\bibfnamefont {L.}~\bibnamefont {Nahon}}, \
  and\ \bibinfo {author} {\bibfnamefont {I.}~\bibnamefont {Powis}},\ }\href
  {\doibase 10.1063/1.1807578} {\bibfield  {journal} {\bibinfo  {journal}
  {Review of Scientific Instruments}\ }\textbf {\bibinfo {volume} {75}},\
  \bibinfo {pages} {4989} (\bibinfo {year} {2004})}\BibitemShut {NoStop}%
\bibitem [{\citenamefont {Garcia}\ \emph {et~al.}(2005)\citenamefont {Garcia},
  \citenamefont {Nahon}, \citenamefont {Harding}, \citenamefont {Mikajlo},\
  and\ \citenamefont {Powis}}]{garcia_refocusing_2005}%
  \BibitemOpen
  \bibfield  {author} {\bibinfo {author} {\bibfnamefont {G.~A.}\ \bibnamefont
  {Garcia}}, \bibinfo {author} {\bibfnamefont {L.}~\bibnamefont {Nahon}},
  \bibinfo {author} {\bibfnamefont {C.~J.}\ \bibnamefont {Harding}}, \bibinfo
  {author} {\bibfnamefont {E.~A.}\ \bibnamefont {Mikajlo}}, \ and\ \bibinfo
  {author} {\bibfnamefont {I.}~\bibnamefont {Powis}},\ }\href {\doibase
  10.1063/1.1900646} {\bibfield  {journal} {\bibinfo  {journal} {Review of
  Scientific Instruments}\ }\textbf {\bibinfo {volume} {76}},\ \bibinfo {pages}
  {053302} (\bibinfo {year} {2005})}\BibitemShut {NoStop}%
\bibitem [{\citenamefont {Rolles}\ \emph {et~al.}(2007)\citenamefont {Rolles},
  \citenamefont {Pe{\v s}i{\'c}}, \citenamefont {Perri}, \citenamefont
  {Bilodeau}, \citenamefont {Ackerman}, \citenamefont {Rude}, \citenamefont
  {Kilcoyne}, \citenamefont {Bozek},\ and\ \citenamefont
  {Berrah}}]{rolles_velocity_2007}%
  \BibitemOpen
  \bibfield  {author} {\bibinfo {author} {\bibfnamefont {D.}~\bibnamefont
  {Rolles}}, \bibinfo {author} {\bibfnamefont {Z.}~\bibnamefont {Pe{\v
  s}i{\'c}}}, \bibinfo {author} {\bibfnamefont {M.}~\bibnamefont {Perri}},
  \bibinfo {author} {\bibfnamefont {R.}~\bibnamefont {Bilodeau}}, \bibinfo
  {author} {\bibfnamefont {G.}~\bibnamefont {Ackerman}}, \bibinfo {author}
  {\bibfnamefont {B.}~\bibnamefont {Rude}}, \bibinfo {author} {\bibfnamefont
  {A.}~\bibnamefont {Kilcoyne}}, \bibinfo {author} {\bibfnamefont
  {J.}~\bibnamefont {Bozek}}, \ and\ \bibinfo {author} {\bibfnamefont
  {N.}~\bibnamefont {Berrah}},\ }\href {\doibase 10.1016/j.nimb.2007.04.186}
  {\bibfield  {journal} {\bibinfo  {journal} {Nuclear Instruments and Methods
  in Physics Research Section B: Beam Interactions with Materials and Atoms}\
  }\textbf {\bibinfo {volume} {261}},\ \bibinfo {pages} {170} (\bibinfo {year}
  {2007})}\BibitemShut {NoStop}%
\bibitem [{\citenamefont {Pe{\v s}i{\'c}}\ \emph {et~al.}(2007)\citenamefont
  {Pe{\v s}i{\'c}}, \citenamefont {Rolles}, \citenamefont {Perri},
  \citenamefont {Bilodeau}, \citenamefont {Ackerman}, \citenamefont {Rude},
  \citenamefont {Kilcoyne}, \citenamefont {Bozek},\ and\ \citenamefont
  {Berrah}}]{pesic_velocity_2007}%
  \BibitemOpen
  \bibfield  {author} {\bibinfo {author} {\bibfnamefont {Z.}~\bibnamefont
  {Pe{\v s}i{\'c}}}, \bibinfo {author} {\bibfnamefont {D.}~\bibnamefont
  {Rolles}}, \bibinfo {author} {\bibfnamefont {M.}~\bibnamefont {Perri}},
  \bibinfo {author} {\bibfnamefont {R.}~\bibnamefont {Bilodeau}}, \bibinfo
  {author} {\bibfnamefont {G.}~\bibnamefont {Ackerman}}, \bibinfo {author}
  {\bibfnamefont {B.}~\bibnamefont {Rude}}, \bibinfo {author} {\bibfnamefont
  {A.}~\bibnamefont {Kilcoyne}}, \bibinfo {author} {\bibfnamefont
  {J.}~\bibnamefont {Bozek}}, \ and\ \bibinfo {author} {\bibfnamefont
  {N.}~\bibnamefont {Berrah}},\ }\href {\doibase 10.1016/j.elspec.2006.11.046}
  {\bibfield  {journal} {\bibinfo  {journal} {Journal of Electron Spectroscopy
  and Related Phenomena}\ }\textbf {\bibinfo {volume} {155}},\ \bibinfo {pages}
  {155} (\bibinfo {year} {2007})}\BibitemShut {NoStop}%
\bibitem [{\citenamefont {Red}\ \emph {et~al.}(2010)\citenamefont {Red},
  \citenamefont {Juarez}, \citenamefont {Rolles},\ and\ \citenamefont
  {Aguilar}}]{red_exploring_2010}%
  \BibitemOpen
  \bibfield  {author} {\bibinfo {author} {\bibfnamefont {E.~C.}\ \bibnamefont
  {Red}}, \bibinfo {author} {\bibfnamefont {A.~M.}\ \bibnamefont {Juarez}},
  \bibinfo {author} {\bibfnamefont {D.}~\bibnamefont {Rolles}}, \ and\ \bibinfo
  {author} {\bibfnamefont {A.}~\bibnamefont {Aguilar}},\ }\href
  {http://www.redalyc.org/pdf/570/57030352023.pdf} {\bibfield  {journal}
  {\bibinfo  {journal} {Revista Mexicana de Fisica S}\ }\textbf {\bibinfo
  {volume} {56}},\ \bibinfo {pages} {100} (\bibinfo {year} {2010})}\BibitemShut
  {NoStop}%
\bibitem [{\citenamefont {Bodi}\ \emph {et~al.}(2009)\citenamefont {Bodi},
  \citenamefont {Johnson}, \citenamefont {Gerber}, \citenamefont {Gengeliczki},
  \citenamefont {SztÃ¡ray},\ and\ \citenamefont {Baer}}]{Bodi2009}%
  \BibitemOpen
  \bibfield  {author} {\bibinfo {author} {\bibfnamefont {A.}~\bibnamefont
  {Bodi}}, \bibinfo {author} {\bibfnamefont {M.}~\bibnamefont {Johnson}},
  \bibinfo {author} {\bibfnamefont {T.}~\bibnamefont {Gerber}}, \bibinfo
  {author} {\bibfnamefont {Z.}~\bibnamefont {Gengeliczki}}, \bibinfo {author}
  {\bibfnamefont {B.}~\bibnamefont {SztÃ¡ray}}, \ and\ \bibinfo {author}
  {\bibfnamefont {T.}~\bibnamefont {Baer}},\ }\href {\doibase
  10.1063/1.3082016} {\bibfield  {journal} {\bibinfo  {journal} {Review of
  Scientific Instruments}\ }\textbf {\bibinfo {volume} {80}},\ \bibinfo {pages}
  {034101} (\bibinfo {year} {2009})}\BibitemShut {NoStop}%
\bibitem [{\citenamefont {O'Keeffe}\ \emph {et~al.}(2011)\citenamefont
  {O'Keeffe}, \citenamefont {Bolognesi}, \citenamefont {Coreno}, \citenamefont
  {Moise}, \citenamefont {Richter}, \citenamefont {Cautero}, \citenamefont
  {Stebel}, \citenamefont {Sergo}, \citenamefont {Pravica}, \citenamefont
  {Ovcharenko},\ and\ \citenamefont {Avaldi}}]{POKeeffe2011}%
  \BibitemOpen
  \bibfield  {author} {\bibinfo {author} {\bibfnamefont {P.}~\bibnamefont
  {O'Keeffe}}, \bibinfo {author} {\bibfnamefont {P.}~\bibnamefont {Bolognesi}},
  \bibinfo {author} {\bibfnamefont {M.}~\bibnamefont {Coreno}}, \bibinfo
  {author} {\bibfnamefont {A.}~\bibnamefont {Moise}}, \bibinfo {author}
  {\bibfnamefont {R.}~\bibnamefont {Richter}}, \bibinfo {author} {\bibfnamefont
  {G.}~\bibnamefont {Cautero}}, \bibinfo {author} {\bibfnamefont
  {L.}~\bibnamefont {Stebel}}, \bibinfo {author} {\bibfnamefont
  {R.}~\bibnamefont {Sergo}}, \bibinfo {author} {\bibfnamefont
  {L.}~\bibnamefont {Pravica}}, \bibinfo {author} {\bibfnamefont
  {Y.}~\bibnamefont {Ovcharenko}}, \ and\ \bibinfo {author} {\bibfnamefont
  {L.}~\bibnamefont {Avaldi}},\ }\href {\doibase 10.1063/1.3563723} {\bibfield
  {journal} {\bibinfo  {journal} {Review of Scientific Instruments}\ }\textbf
  {\bibinfo {volume} {82}},\ \bibinfo {pages} {033109} (\bibinfo {year}
  {2011})}\BibitemShut {NoStop}%
\bibitem [{\citenamefont {Str\"uder}\ \emph {et~al.}(2010)\citenamefont
  {Str\"uder}, \citenamefont {Epp}, \citenamefont {Rolles}, \citenamefont
  {Hartmann}, \citenamefont {Holl}, \citenamefont {Lutz}, \citenamefont
  {Soltau}, \citenamefont {Eckart}, \citenamefont {Reich}, \citenamefont
  {Heinzinger},\ and\ \citenamefont {et~al.}}]{Strueder2010}%
  \BibitemOpen
  \bibfield  {author} {\bibinfo {author} {\bibfnamefont {L.}~\bibnamefont
  {Str\"uder}}, \bibinfo {author} {\bibfnamefont {S.}~\bibnamefont {Epp}},
  \bibinfo {author} {\bibfnamefont {D.}~\bibnamefont {Rolles}}, \bibinfo
  {author} {\bibfnamefont {R.}~\bibnamefont {Hartmann}}, \bibinfo {author}
  {\bibfnamefont {P.}~\bibnamefont {Holl}}, \bibinfo {author} {\bibfnamefont
  {G.}~\bibnamefont {Lutz}}, \bibinfo {author} {\bibfnamefont {H.}~\bibnamefont
  {Soltau}}, \bibinfo {author} {\bibfnamefont {R.}~\bibnamefont {Eckart}},
  \bibinfo {author} {\bibfnamefont {C.}~\bibnamefont {Reich}}, \bibinfo
  {author} {\bibfnamefont {K.}~\bibnamefont {Heinzinger}}, \ and\ \bibinfo
  {author} {\bibnamefont {et~al.}},\ }\href {\doibase
  10.1016/j.nima.2009.12.053} {\bibfield  {journal} {\bibinfo  {journal}
  {Nuclear Instruments and Methods in Physics Research A}\ }\textbf {\bibinfo
  {volume} {614}},\ \bibinfo {pages} {483–496} (\bibinfo {year}
  {2010})}\BibitemShut {NoStop}%
\bibitem [{\citenamefont {Rolles}\ \emph {et~al.}(2014)\citenamefont {Rolles},
  \citenamefont {Boll}, \citenamefont {Adolph}, \citenamefont {Aquila},
  \citenamefont {Bostedt}, \citenamefont {Bozek}, \citenamefont {Chapman},
  \citenamefont {Coffee}, \citenamefont {Coppola}, \citenamefont {Decleva},
  \citenamefont {Delmas}, \citenamefont {Epp}, \citenamefont {Erk},
  \citenamefont {Filsinger}, \citenamefont {Foucar}, \citenamefont {Gumprecht},
  \citenamefont {H{\"o}mke}, \citenamefont {Gorkhover}, \citenamefont
  {Holmegaard}, \citenamefont {Johnsson}, \citenamefont {Kaiser}, \citenamefont
  {Krasniqi}, \citenamefont {K{\"u}hnel}, \citenamefont {Maurer}, \citenamefont
  {Messerschmidt}, \citenamefont {Moshammer}, \citenamefont {Quevedo},
  \citenamefont {Rajkovic}, \citenamefont {Rouz{\'e}e}, \citenamefont {Rudek},
  \citenamefont {Schlichting}, \citenamefont {Schmidt}, \citenamefont {Schorb},
  \citenamefont {Schr{\"o}ter}, \citenamefont {Schulz}, \citenamefont
  {Stapelfeldt}, \citenamefont {Stener}, \citenamefont {Stern}, \citenamefont
  {Techert}, \citenamefont {Th{\o}gersen}, \citenamefont {Vrakking},
  \citenamefont {Rudenko}, \citenamefont {K{\"u}pper},\ and\ \citenamefont
  {Ullrich}}]{rolles_femtosecond_2014}%
  \BibitemOpen
  \bibfield  {author} {\bibinfo {author} {\bibfnamefont {D.}~\bibnamefont
  {Rolles}}, \bibinfo {author} {\bibfnamefont {R.}~\bibnamefont {Boll}},
  \bibinfo {author} {\bibfnamefont {M.}~\bibnamefont {Adolph}}, \bibinfo
  {author} {\bibfnamefont {A.}~\bibnamefont {Aquila}}, \bibinfo {author}
  {\bibfnamefont {C.}~\bibnamefont {Bostedt}}, \bibinfo {author} {\bibfnamefont
  {J.~D.}\ \bibnamefont {Bozek}}, \bibinfo {author} {\bibfnamefont {H.~N.}\
  \bibnamefont {Chapman}}, \bibinfo {author} {\bibfnamefont {R.}~\bibnamefont
  {Coffee}}, \bibinfo {author} {\bibfnamefont {N.}~\bibnamefont {Coppola}},
  \bibinfo {author} {\bibfnamefont {P.}~\bibnamefont {Decleva}}, \bibinfo
  {author} {\bibfnamefont {T.}~\bibnamefont {Delmas}}, \bibinfo {author}
  {\bibfnamefont {S.~W.}\ \bibnamefont {Epp}}, \bibinfo {author} {\bibfnamefont
  {B.}~\bibnamefont {Erk}}, \bibinfo {author} {\bibfnamefont {F.}~\bibnamefont
  {Filsinger}}, \bibinfo {author} {\bibfnamefont {L.}~\bibnamefont {Foucar}},
  \bibinfo {author} {\bibfnamefont {L.}~\bibnamefont {Gumprecht}}, \bibinfo
  {author} {\bibfnamefont {A.}~\bibnamefont {H{\"o}mke}}, \bibinfo {author}
  {\bibfnamefont {T.}~\bibnamefont {Gorkhover}}, \bibinfo {author}
  {\bibfnamefont {L.}~\bibnamefont {Holmegaard}}, \bibinfo {author}
  {\bibfnamefont {P.}~\bibnamefont {Johnsson}}, \bibinfo {author}
  {\bibfnamefont {C.}~\bibnamefont {Kaiser}}, \bibinfo {author} {\bibfnamefont
  {F.}~\bibnamefont {Krasniqi}}, \bibinfo {author} {\bibfnamefont {K.-U.}\
  \bibnamefont {K{\"u}hnel}}, \bibinfo {author} {\bibfnamefont
  {J.}~\bibnamefont {Maurer}}, \bibinfo {author} {\bibfnamefont
  {M.}~\bibnamefont {Messerschmidt}}, \bibinfo {author} {\bibfnamefont
  {R.}~\bibnamefont {Moshammer}}, \bibinfo {author} {\bibfnamefont
  {W.}~\bibnamefont {Quevedo}}, \bibinfo {author} {\bibfnamefont
  {I.}~\bibnamefont {Rajkovic}}, \bibinfo {author} {\bibfnamefont
  {A.}~\bibnamefont {Rouz{\'e}e}}, \bibinfo {author} {\bibfnamefont
  {B.}~\bibnamefont {Rudek}}, \bibinfo {author} {\bibfnamefont
  {I.}~\bibnamefont {Schlichting}}, \bibinfo {author} {\bibfnamefont
  {C.}~\bibnamefont {Schmidt}}, \bibinfo {author} {\bibfnamefont
  {S.}~\bibnamefont {Schorb}}, \bibinfo {author} {\bibfnamefont {C.~D.}\
  \bibnamefont {Schr{\"o}ter}}, \bibinfo {author} {\bibfnamefont
  {J.}~\bibnamefont {Schulz}}, \bibinfo {author} {\bibfnamefont
  {H.}~\bibnamefont {Stapelfeldt}}, \bibinfo {author} {\bibfnamefont
  {M.}~\bibnamefont {Stener}}, \bibinfo {author} {\bibfnamefont
  {S.}~\bibnamefont {Stern}}, \bibinfo {author} {\bibfnamefont
  {S.}~\bibnamefont {Techert}}, \bibinfo {author} {\bibfnamefont
  {J.}~\bibnamefont {Th{\o}gersen}}, \bibinfo {author} {\bibfnamefont
  {M.~J.~J.}\ \bibnamefont {Vrakking}}, \bibinfo {author} {\bibfnamefont
  {A.}~\bibnamefont {Rudenko}}, \bibinfo {author} {\bibfnamefont
  {J.}~\bibnamefont {K{\"u}pper}}, \ and\ \bibinfo {author} {\bibfnamefont
  {J.}~\bibnamefont {Ullrich}},\ }\href {\doibase
  10.1088/0953-4075/47/12/124035} {\bibfield  {journal} {\bibinfo  {journal}
  {Journal of Physics B: Atomic, Molecular and Optical Physics}\ }\textbf
  {\bibinfo {volume} {47}},\ \bibinfo {pages} {124035} (\bibinfo {year}
  {2014})}\BibitemShut {NoStop}%
\bibitem [{\citenamefont {Takahashi}, \citenamefont {Cave},\ and\ \citenamefont
  {Eland}(2000)}]{Takahashi2000}%
  \BibitemOpen
  \bibfield  {author} {\bibinfo {author} {\bibfnamefont {M.}~\bibnamefont
  {Takahashi}}, \bibinfo {author} {\bibfnamefont {J.~P.}\ \bibnamefont {Cave}},
  \ and\ \bibinfo {author} {\bibfnamefont {J.~H.~D.}\ \bibnamefont {Eland}},\
  }\href {\doibase 10.1063/1.1150460} {\bibfield  {journal} {\bibinfo
  {journal} {Review of Scientific Instruments}\ }\textbf {\bibinfo {volume}
  {71}},\ \bibinfo {pages} {1337} (\bibinfo {year} {2000})}\BibitemShut
  {NoStop}%
\bibitem [{\citenamefont {Vredenborg}, \citenamefont {Roeterdink},\ and\
  \citenamefont {Janssen}(2008{\natexlab{a}})}]{Vredenborg2008}%
  \BibitemOpen
  \bibfield  {author} {\bibinfo {author} {\bibfnamefont {A.}~\bibnamefont
  {Vredenborg}}, \bibinfo {author} {\bibfnamefont {W.~G.}\ \bibnamefont
  {Roeterdink}}, \ and\ \bibinfo {author} {\bibfnamefont {M.~H.~M.}\
  \bibnamefont {Janssen}},\ }\href {\doibase 10.1063/1.2949142} {\bibfield
  {journal} {\bibinfo  {journal} {Review of Scientific Instruments}\ }\textbf
  {\bibinfo {volume} {79}},\ \bibinfo {pages} {063108} (\bibinfo {year}
  {2008}{\natexlab{a}})}\BibitemShut {NoStop}%
\bibitem [{\citenamefont {Garcia}\ \emph {et~al.}(2013)\citenamefont {Garcia},
  \citenamefont {Cunha~de Miranda}, \citenamefont {Tia}, \citenamefont {Daly},\
  and\ \citenamefont {Nahon}}]{garcia_delicious_2013}%
  \BibitemOpen
  \bibfield  {author} {\bibinfo {author} {\bibfnamefont {G.~A.}\ \bibnamefont
  {Garcia}}, \bibinfo {author} {\bibfnamefont {B.~K.}\ \bibnamefont {Cunha~de
  Miranda}}, \bibinfo {author} {\bibfnamefont {M.}~\bibnamefont {Tia}},
  \bibinfo {author} {\bibfnamefont {S.}~\bibnamefont {Daly}}, \ and\ \bibinfo
  {author} {\bibfnamefont {L.}~\bibnamefont {Nahon}},\ }\href {\doibase
  10.1063/1.4807751} {\bibfield  {journal} {\bibinfo  {journal} {Review of
  Scientific Instruments}\ }\textbf {\bibinfo {volume} {84}},\ \bibinfo {pages}
  {053112} (\bibinfo {year} {2013})}\BibitemShut {NoStop}%
\bibitem [{\citenamefont {Bomme}\ \emph {et~al.}(2013)\citenamefont {Bomme},
  \citenamefont {Guillemin}, \citenamefont {Marin}, \citenamefont {Journel},
  \citenamefont {Marchenko}, \citenamefont {Dowek}, \citenamefont {Trcera},
  \citenamefont {Pilette}, \citenamefont {Avila}, \citenamefont {Ringuenet},
  \citenamefont {Kushawaha},\ and\ \citenamefont {Simon}}]{bomme_double_2013}%
  \BibitemOpen
  \bibfield  {author} {\bibinfo {author} {\bibfnamefont {C.}~\bibnamefont
  {Bomme}}, \bibinfo {author} {\bibfnamefont {R.}~\bibnamefont {Guillemin}},
  \bibinfo {author} {\bibfnamefont {T.}~\bibnamefont {Marin}}, \bibinfo
  {author} {\bibfnamefont {L.}~\bibnamefont {Journel}}, \bibinfo {author}
  {\bibfnamefont {T.}~\bibnamefont {Marchenko}}, \bibinfo {author}
  {\bibfnamefont {D.}~\bibnamefont {Dowek}}, \bibinfo {author} {\bibfnamefont
  {N.}~\bibnamefont {Trcera}}, \bibinfo {author} {\bibfnamefont
  {B.}~\bibnamefont {Pilette}}, \bibinfo {author} {\bibfnamefont
  {A.}~\bibnamefont {Avila}}, \bibinfo {author} {\bibfnamefont
  {H.}~\bibnamefont {Ringuenet}}, \bibinfo {author} {\bibfnamefont {R.~K.}\
  \bibnamefont {Kushawaha}}, \ and\ \bibinfo {author} {\bibfnamefont
  {M.}~\bibnamefont {Simon}},\ }\href {\doibase 10.1063/1.4824194} {\bibfield
  {journal} {\bibinfo  {journal} {Review of Scientific Instruments}\ }\textbf
  {\bibinfo {volume} {84}},\ \bibinfo {pages} {103104} (\bibinfo {year}
  {2013})}\BibitemShut {NoStop}%
\bibitem [{\citenamefont {Szt{\'a}ray}\ \emph {et~al.}(2017)\citenamefont
  {Szt{\'a}ray}, \citenamefont {Voronova}, \citenamefont {Torma}, \citenamefont
  {Covert}, \citenamefont {Bodi}, \citenamefont {Hemberger}, \citenamefont
  {Gerber},\ and\ \citenamefont {Osborn}}]{sztaray_crf-pepico:_2017}%
  \BibitemOpen
  \bibfield  {author} {\bibinfo {author} {\bibfnamefont {B.}~\bibnamefont
  {Szt{\'a}ray}}, \bibinfo {author} {\bibfnamefont {K.}~\bibnamefont
  {Voronova}}, \bibinfo {author} {\bibfnamefont {K.~G.}\ \bibnamefont {Torma}},
  \bibinfo {author} {\bibfnamefont {K.~J.}\ \bibnamefont {Covert}}, \bibinfo
  {author} {\bibfnamefont {A.}~\bibnamefont {Bodi}}, \bibinfo {author}
  {\bibfnamefont {P.}~\bibnamefont {Hemberger}}, \bibinfo {author}
  {\bibfnamefont {T.}~\bibnamefont {Gerber}}, \ and\ \bibinfo {author}
  {\bibfnamefont {D.~L.}\ \bibnamefont {Osborn}},\ }\href {\doibase
  10.1063/1.4984304} {\bibfield  {journal} {\bibinfo  {journal} {The Journal of
  Chemical Physics}\ }\textbf {\bibinfo {volume} {147}},\ \bibinfo {pages}
  {013944} (\bibinfo {year} {2017})}\BibitemShut {NoStop}%
\bibitem [{\citenamefont {Hosaka}\ \emph {et~al.}(2006)\citenamefont {Hosaka},
  \citenamefont {i.~Adachi}, \citenamefont {Golovin}, \citenamefont
  {Takahashi}, \citenamefont {Watanabe},\ and\ \citenamefont
  {Yagishita}}]{Hosaka2006}%
  \BibitemOpen
  \bibfield  {author} {\bibinfo {author} {\bibfnamefont {K.}~\bibnamefont
  {Hosaka}}, \bibinfo {author} {\bibfnamefont {J.}~\bibnamefont {i.~Adachi}},
  \bibinfo {author} {\bibfnamefont {A.~V.}\ \bibnamefont {Golovin}}, \bibinfo
  {author} {\bibfnamefont {M.}~\bibnamefont {Takahashi}}, \bibinfo {author}
  {\bibfnamefont {N.}~\bibnamefont {Watanabe}}, \ and\ \bibinfo {author}
  {\bibfnamefont {A.}~\bibnamefont {Yagishita}},\ }\href
  {http://stacks.iop.org/1347-4065/45/i=3R/a=1841} {\bibfield  {journal}
  {\bibinfo  {journal} {Japanese Journal of Applied Physics}\ }\textbf
  {\bibinfo {volume} {45}},\ \bibinfo {pages} {1841} (\bibinfo {year}
  {2006})}\BibitemShut {NoStop}%
\bibitem [{\citenamefont {Tang}\ \emph {et~al.}(2009)\citenamefont {Tang},
  \citenamefont {Zhou}, \citenamefont {Niu}, \citenamefont {Liu}, \citenamefont
  {Sun}, \citenamefont {Shan}, \citenamefont {Liu},\ and\ \citenamefont
  {Sheng}}]{Xiaofeng2009}%
  \BibitemOpen
  \bibfield  {author} {\bibinfo {author} {\bibfnamefont {X.}~\bibnamefont
  {Tang}}, \bibinfo {author} {\bibfnamefont {X.}~\bibnamefont {Zhou}}, \bibinfo
  {author} {\bibfnamefont {M.}~\bibnamefont {Niu}}, \bibinfo {author}
  {\bibfnamefont {S.}~\bibnamefont {Liu}}, \bibinfo {author} {\bibfnamefont
  {J.}~\bibnamefont {Sun}}, \bibinfo {author} {\bibfnamefont {X.}~\bibnamefont
  {Shan}}, \bibinfo {author} {\bibfnamefont {F.}~\bibnamefont {Liu}}, \ and\
  \bibinfo {author} {\bibfnamefont {L.}~\bibnamefont {Sheng}},\ }\href
  {\doibase 10.1063/1.3250872} {\bibfield  {journal} {\bibinfo  {journal}
  {Review of Scientific Instruments}\ }\textbf {\bibinfo {volume} {80}},\
  \bibinfo {pages} {113101} (\bibinfo {year} {2009})}\BibitemShut {NoStop}%
\bibitem [{\citenamefont {Bodi}\ \emph {et~al.}(2012)\citenamefont {Bodi},
  \citenamefont {Hemberger}, \citenamefont {Gerber},\ and\ \citenamefont
  {SztÃ¡ray}}]{bodi2012}%
  \BibitemOpen
  \bibfield  {author} {\bibinfo {author} {\bibfnamefont {A.}~\bibnamefont
  {Bodi}}, \bibinfo {author} {\bibfnamefont {P.}~\bibnamefont {Hemberger}},
  \bibinfo {author} {\bibfnamefont {T.}~\bibnamefont {Gerber}}, \ and\ \bibinfo
  {author} {\bibfnamefont {B.}~\bibnamefont {SztÃ¡ray}},\ }\href {\doibase
  10.1063/1.4742769} {\bibfield  {journal} {\bibinfo  {journal} {Review of
  Scientific Instruments}\ }\textbf {\bibinfo {volume} {83}},\ \bibinfo {pages}
  {083105} (\bibinfo {year} {2012})}\BibitemShut {NoStop}%
\bibitem [{\citenamefont {Ablikim}\ \emph {et~al.}(2016)\citenamefont
  {Ablikim}, \citenamefont {Bomme}, \citenamefont {Xiong}, \citenamefont
  {Savelyev}, \citenamefont {Obaid}, \citenamefont {Kaderiya}, \citenamefont
  {Augustin}, \citenamefont {Schnorr}, \citenamefont {Dumitriu}, \citenamefont
  {Osipov}, \citenamefont {Bilodeau}, \citenamefont {Kilcoyne}, \citenamefont
  {Kumarappan}, \citenamefont {Rudenko}, \citenamefont {Berrah},\ and\
  \citenamefont {Rolles}}]{Ablikim2016}%
  \BibitemOpen
  \bibfield  {author} {\bibinfo {author} {\bibfnamefont {U.}~\bibnamefont
  {Ablikim}}, \bibinfo {author} {\bibfnamefont {C.}~\bibnamefont {Bomme}},
  \bibinfo {author} {\bibfnamefont {H.}~\bibnamefont {Xiong}}, \bibinfo
  {author} {\bibfnamefont {E.}~\bibnamefont {Savelyev}}, \bibinfo {author}
  {\bibfnamefont {R.}~\bibnamefont {Obaid}}, \bibinfo {author} {\bibfnamefont
  {B.}~\bibnamefont {Kaderiya}}, \bibinfo {author} {\bibfnamefont
  {S.}~\bibnamefont {Augustin}}, \bibinfo {author} {\bibfnamefont
  {K.}~\bibnamefont {Schnorr}}, \bibinfo {author} {\bibfnamefont
  {I.}~\bibnamefont {Dumitriu}}, \bibinfo {author} {\bibfnamefont
  {T.}~\bibnamefont {Osipov}}, \bibinfo {author} {\bibfnamefont
  {R.}~\bibnamefont {Bilodeau}}, \bibinfo {author} {\bibfnamefont
  {D.}~\bibnamefont {Kilcoyne}}, \bibinfo {author} {\bibfnamefont
  {V.}~\bibnamefont {Kumarappan}}, \bibinfo {author} {\bibfnamefont
  {A.}~\bibnamefont {Rudenko}}, \bibinfo {author} {\bibfnamefont
  {N.}~\bibnamefont {Berrah}}, \ and\ \bibinfo {author} {\bibfnamefont
  {D.}~\bibnamefont {Rolles}},\ }\href {\doibase 10.1038/srep38202} {\bibfield
  {journal} {\bibinfo  {journal} {Scientific Reports}\ }\textbf {\bibinfo
  {volume} {6}},\ \bibinfo {pages} {38202} (\bibinfo {year}
  {2016})}\BibitemShut {NoStop}%
\bibitem [{\citenamefont {Ablikim}\ \emph {et~al.}(2017)\citenamefont
  {Ablikim}, \citenamefont {Bomme}, \citenamefont {Savelyev}, \citenamefont
  {Xiong}, \citenamefont {Kushawaha}, \citenamefont {Boll}, \citenamefont
  {Amini}, \citenamefont {Osipov}, \citenamefont {Kilcoyne}, \citenamefont
  {Rudenko}, \citenamefont {Berrah},\ and\ \citenamefont
  {Rolles}}]{Ablikim2017}%
  \BibitemOpen
  \bibfield  {author} {\bibinfo {author} {\bibfnamefont {U.}~\bibnamefont
  {Ablikim}}, \bibinfo {author} {\bibfnamefont {C.}~\bibnamefont {Bomme}},
  \bibinfo {author} {\bibfnamefont {E.}~\bibnamefont {Savelyev}}, \bibinfo
  {author} {\bibfnamefont {H.}~\bibnamefont {Xiong}}, \bibinfo {author}
  {\bibfnamefont {R.}~\bibnamefont {Kushawaha}}, \bibinfo {author}
  {\bibfnamefont {R.}~\bibnamefont {Boll}}, \bibinfo {author} {\bibfnamefont
  {K.}~\bibnamefont {Amini}}, \bibinfo {author} {\bibfnamefont
  {T.}~\bibnamefont {Osipov}}, \bibinfo {author} {\bibfnamefont
  {D.}~\bibnamefont {Kilcoyne}}, \bibinfo {author} {\bibfnamefont
  {A.}~\bibnamefont {Rudenko}}, \bibinfo {author} {\bibfnamefont
  {N.}~\bibnamefont {Berrah}}, \ and\ \bibinfo {author} {\bibfnamefont
  {D.}~\bibnamefont {Rolles}},\ }\href {\doibase 10.1039/C7CP01379E} {\bibfield
   {journal} {\bibinfo  {journal} {Physical Chemistry Chemical Physics}\
  }\textbf {\bibinfo {volume} {19}},\ \bibinfo {pages} {13419–13431}
  (\bibinfo {year} {2017})}\BibitemShut {NoStop}%
\bibitem [{\citenamefont {Xiong}\ \emph {et~al.}(2017)\citenamefont {Xiong},
  \citenamefont {Obaid}, \citenamefont {Fang}, \citenamefont {Bomme},
  \citenamefont {Kling}, \citenamefont {Ablikim}, \citenamefont {Petrovic},
  \citenamefont {Liekhus-Schmaltz}, \citenamefont {Li}, \citenamefont
  {Bilodeau}, \citenamefont {Wolf}, \citenamefont {Osipov}, \citenamefont
  {Rolles},\ and\ \citenamefont {Berrah}}]{HuiXiong2017}%
  \BibitemOpen
  \bibfield  {author} {\bibinfo {author} {\bibfnamefont {H.}~\bibnamefont
  {Xiong}}, \bibinfo {author} {\bibfnamefont {R.}~\bibnamefont {Obaid}},
  \bibinfo {author} {\bibfnamefont {L.}~\bibnamefont {Fang}}, \bibinfo {author}
  {\bibfnamefont {C.}~\bibnamefont {Bomme}}, \bibinfo {author} {\bibfnamefont
  {N.~G.}\ \bibnamefont {Kling}}, \bibinfo {author} {\bibfnamefont
  {U.}~\bibnamefont {Ablikim}}, \bibinfo {author} {\bibfnamefont
  {V.}~\bibnamefont {Petrovic}}, \bibinfo {author} {\bibfnamefont {C.~E.}\
  \bibnamefont {Liekhus-Schmaltz}}, \bibinfo {author} {\bibfnamefont
  {H.}~\bibnamefont {Li}}, \bibinfo {author} {\bibfnamefont {R.~C.}\
  \bibnamefont {Bilodeau}}, \bibinfo {author} {\bibfnamefont {T.}~\bibnamefont
  {Wolf}}, \bibinfo {author} {\bibfnamefont {T.}~\bibnamefont {Osipov}},
  \bibinfo {author} {\bibfnamefont {D.}~\bibnamefont {Rolles}}, \ and\ \bibinfo
  {author} {\bibfnamefont {N.}~\bibnamefont {Berrah}},\ }\href {\doibase
  10.1103/PhysRevA.96.033408} {\bibfield  {journal} {\bibinfo  {journal} {Phys.
  Rev. A}\ }\textbf {\bibinfo {volume} {96}},\ \bibinfo {pages} {033408}
  (\bibinfo {year} {2017})}\BibitemShut {NoStop}%
\bibitem [{\citenamefont {Lehmann}, \citenamefont {Ram},\ and\ \citenamefont
  {Janssen}(2012)}]{Lehmann2012}%
  \BibitemOpen
  \bibfield  {author} {\bibinfo {author} {\bibfnamefont {C.~S.}\ \bibnamefont
  {Lehmann}}, \bibinfo {author} {\bibfnamefont {N.~B.}\ \bibnamefont {Ram}}, \
  and\ \bibinfo {author} {\bibfnamefont {M.~H.~M.}\ \bibnamefont {Janssen}},\
  }\href {\doibase 10.1063/1.4749843} {\bibfield  {journal} {\bibinfo
  {journal} {Review of Scientific Instruments}\ }\textbf {\bibinfo {volume}
  {83}},\ \bibinfo {pages} {093103} (\bibinfo {year} {2012})}\BibitemShut
  {NoStop}%
\bibitem [{\citenamefont {Zhao}, \citenamefont {S\'andor},\ and\ \citenamefont
  {Weinacht}(2017)}]{Zhao2017}%
  \BibitemOpen
  \bibfield  {author} {\bibinfo {author} {\bibfnamefont {A.}~\bibnamefont
  {Zhao}}, \bibinfo {author} {\bibfnamefont {P.}~\bibnamefont {S\'andor}}, \
  and\ \bibinfo {author} {\bibfnamefont {T.}~\bibnamefont {Weinacht}},\ }\href
  {\doibase 10.1063/1.4981917} {\bibfield  {journal} {\bibinfo  {journal} {The
  Journal of Chemical Physics}\ }\textbf {\bibinfo {volume} {147}},\ \bibinfo
  {pages} {013922} (\bibinfo {year} {2017})}\BibitemShut {NoStop}%
\bibitem [{\citenamefont {Fan}\ \emph {et~al.}(2017)\citenamefont {Fan},
  \citenamefont {Lee}, \citenamefont {Tu}, \citenamefont {Mignolet},
  \citenamefont {Couch}, \citenamefont {Dorney}, \citenamefont {Nguyen},
  \citenamefont {Wooldridge}, \citenamefont {Murnane}, \citenamefont
  {Remacle},\ and\ \citenamefont {et~al.}}]{Fan2017}%
  \BibitemOpen
  \bibfield  {author} {\bibinfo {author} {\bibfnamefont {L.}~\bibnamefont
  {Fan}}, \bibinfo {author} {\bibfnamefont {S.~K.}\ \bibnamefont {Lee}},
  \bibinfo {author} {\bibfnamefont {Y.-J.}\ \bibnamefont {Tu}}, \bibinfo
  {author} {\bibfnamefont {B.}~\bibnamefont {Mignolet}}, \bibinfo {author}
  {\bibfnamefont {D.}~\bibnamefont {Couch}}, \bibinfo {author} {\bibfnamefont
  {K.}~\bibnamefont {Dorney}}, \bibinfo {author} {\bibfnamefont
  {Q.}~\bibnamefont {Nguyen}}, \bibinfo {author} {\bibfnamefont
  {L.}~\bibnamefont {Wooldridge}}, \bibinfo {author} {\bibfnamefont
  {M.}~\bibnamefont {Murnane}}, \bibinfo {author} {\bibfnamefont
  {F.}~\bibnamefont {Remacle}}, \ and\ \bibinfo {author} {\bibnamefont
  {et~al.}},\ }\href {\doibase 10.1063/1.4981526} {\bibfield  {journal}
  {\bibinfo  {journal} {The Journal of Chemical Physics}\ }\textbf {\bibinfo
  {volume} {147}},\ \bibinfo {pages} {013920} (\bibinfo {year}
  {2017})}\BibitemShut {NoStop}%
\bibitem [{\citenamefont {Kaindl}\ \emph {et~al.}(1995)\citenamefont {Kaindl},
  \citenamefont {Schulz}, \citenamefont {Heimann}, \citenamefont {Bozek},\ and\
  \citenamefont {Schlachter}}]{kaindl_ultrahigh_1995}%
  \BibitemOpen
  \bibfield  {author} {\bibinfo {author} {\bibfnamefont {G.}~\bibnamefont
  {Kaindl}}, \bibinfo {author} {\bibfnamefont {K.}~\bibnamefont {Schulz}},
  \bibinfo {author} {\bibfnamefont {P.}~\bibnamefont {Heimann}}, \bibinfo
  {author} {\bibfnamefont {J.}~\bibnamefont {Bozek}}, \ and\ \bibinfo {author}
  {\bibfnamefont {A.}~\bibnamefont {Schlachter}},\ }\href {\doibase
  10.1080/08940889508602839} {\bibfield  {journal} {\bibinfo  {journal}
  {Synchrotron Radiation News}\ }\textbf {\bibinfo {volume} {8}},\ \bibinfo
  {pages} {29} (\bibinfo {year} {1995})}\BibitemShut {NoStop}%
\bibitem [{\citenamefont {Warwick}\ \emph {et~al.}(1995)\citenamefont
  {Warwick}, \citenamefont {Andresen}, \citenamefont {Portmann},\ and\
  \citenamefont {Jackson}}]{warwick_performance_1995}%
  \BibitemOpen
  \bibfield  {author} {\bibinfo {author} {\bibfnamefont {T.}~\bibnamefont
  {Warwick}}, \bibinfo {author} {\bibfnamefont {N.}~\bibnamefont {Andresen}},
  \bibinfo {author} {\bibfnamefont {G.}~\bibnamefont {Portmann}}, \ and\
  \bibinfo {author} {\bibfnamefont {A.}~\bibnamefont {Jackson}},\ }\href
  {\doibase 10.1063/1.1145777} {\bibfield  {journal} {\bibinfo  {journal}
  {Review of Scientific Instruments}\ }\textbf {\bibinfo {volume} {66}},\
  \bibinfo {pages} {1984} (\bibinfo {year} {1995})}\BibitemShut {NoStop}%
\bibitem [{\citenamefont {Vredenborg}, \citenamefont {Roeterdink},\ and\
  \citenamefont
  {Janssen}(2008{\natexlab{b}})}]{vredenborg_photoelectron-photoion_2008}%
  \BibitemOpen
  \bibfield  {author} {\bibinfo {author} {\bibfnamefont {A.}~\bibnamefont
  {Vredenborg}}, \bibinfo {author} {\bibfnamefont {W.~G.}\ \bibnamefont
  {Roeterdink}}, \ and\ \bibinfo {author} {\bibfnamefont {M.~H.~M.}\
  \bibnamefont {Janssen}},\ }\href {\doibase 10.1063/1.2949142} {\bibfield
  {journal} {\bibinfo  {journal} {Review of Scientific Instruments}\ }\textbf
  {\bibinfo {volume} {79}},\ \bibinfo {pages} {063108} (\bibinfo {year}
  {2008}{\natexlab{b}})}\BibitemShut {NoStop}%
\bibitem [{\citenamefont {Hickstein}\ \emph {et~al.}()\citenamefont
  {Hickstein}, \citenamefont {Yurchak}, \citenamefont {Das}, \citenamefont
  {Shih},\ and\ \citenamefont {Stephen}}]{hickstein_pyabel_nodate}%
  \BibitemOpen
  \bibfield  {author} {\bibinfo {author} {\bibfnamefont {D.}~\bibnamefont
  {Hickstein}}, \bibinfo {author} {\bibfnamefont {R.}~\bibnamefont {Yurchak}},
  \bibinfo {author} {\bibfnamefont {D.}~\bibnamefont {Das}}, \bibinfo {author}
  {\bibfnamefont {C.-Y.}\ \bibnamefont {Shih}}, \ and\ \bibinfo {author}
  {\bibfnamefont {T.~G.}\ \bibnamefont {Stephen}},\ }\href
  {https://zenodo.org/record/47423#.WpnV7ejwaM8} {\enquote {\bibinfo {title}
  {{PyAbel} (v0.7): {A} {Python} {Package} for {Abel} {Transforms}},}\
  }\BibitemShut {NoStop}%
\bibitem [{\citenamefont {Gascooke}, \citenamefont {Gibson},\ and\
  \citenamefont {Lawrance}(2017)}]{Gascooke2017}%
  \BibitemOpen
  \bibfield  {author} {\bibinfo {author} {\bibfnamefont {J.~R.}\ \bibnamefont
  {Gascooke}}, \bibinfo {author} {\bibfnamefont {S.~T.}\ \bibnamefont
  {Gibson}}, \ and\ \bibinfo {author} {\bibfnamefont {W.~D.}\ \bibnamefont
  {Lawrance}},\ }\href {\doibase 10.1063/1.4981024} {\bibfield  {journal}
  {\bibinfo  {journal} {The Journal of Chemical Physics}\ }\textbf {\bibinfo
  {volume} {147}},\ \bibinfo {pages} {013924} (\bibinfo {year}
  {2017})}\BibitemShut {NoStop}%
\bibitem [{\citenamefont {Weber}\ \emph {et~al.}(2001)\citenamefont {Weber},
  \citenamefont {Jagutzki}, \citenamefont {Hattass}, \citenamefont {Staudte},
  \citenamefont {Nauert}, \citenamefont {Schmidt}, \citenamefont {Prior},
  \citenamefont {Landers}, \citenamefont {Br{\"a}uning-Demian}, \citenamefont
  {Br{\"a}uning},\ and\ \citenamefont {Garc{\'i}a~de
  Abajo}}]{weber_k-shell_2001}%
  \BibitemOpen
  \bibfield  {author} {\bibinfo {author} {\bibfnamefont {T.}~\bibnamefont
  {Weber}}, \bibinfo {author} {\bibfnamefont {O.}~\bibnamefont {Jagutzki}},
  \bibinfo {author} {\bibfnamefont {M.}~\bibnamefont {Hattass}}, \bibinfo
  {author} {\bibfnamefont {A.}~\bibnamefont {Staudte}}, \bibinfo {author}
  {\bibfnamefont {A.}~\bibnamefont {Nauert}}, \bibinfo {author} {\bibfnamefont
  {L.}~\bibnamefont {Schmidt}}, \bibinfo {author} {\bibfnamefont {M.~H.}\
  \bibnamefont {Prior}}, \bibinfo {author} {\bibfnamefont {A.~L.}\ \bibnamefont
  {Landers}}, \bibinfo {author} {\bibfnamefont {A.}~\bibnamefont
  {Br{\"a}uning-Demian}}, \bibinfo {author} {\bibfnamefont {H.}~\bibnamefont
  {Br{\"a}uning}}, \ and\ \bibinfo {author} {\bibfnamefont {F.}~\bibnamefont
  {Garc{\'i}a~de Abajo}},\ }\href
  {http://iopscience.iop.org/0953-4075/34/18/305} {\bibfield  {journal}
  {\bibinfo  {journal} {Journal of Physics B: Atomic, Molecular and Optical
  Physics}\ }\textbf {\bibinfo {volume} {34}},\ \bibinfo {pages} {3669}
  (\bibinfo {year} {2001})}\BibitemShut {NoStop}%
\bibitem [{\citenamefont {Lundqvist}\ \emph {et~al.}(1996)\citenamefont
  {Lundqvist}, \citenamefont {Edvardsson}, \citenamefont {Baltzer},\ and\
  \citenamefont {Wannberg}}]{lundqvist_doppler-free_1996}%
  \BibitemOpen
  \bibfield  {author} {\bibinfo {author} {\bibfnamefont {M.}~\bibnamefont
  {Lundqvist}}, \bibinfo {author} {\bibfnamefont {D.}~\bibnamefont
  {Edvardsson}}, \bibinfo {author} {\bibfnamefont {P.}~\bibnamefont {Baltzer}},
  \ and\ \bibinfo {author} {\bibfnamefont {B.}~\bibnamefont {Wannberg}},\
  }\href {http://iopscience.iop.org/0953-4075/29/8/013} {\bibfield  {journal}
  {\bibinfo  {journal} {Journal of Physics B: Atomic, Molecular and Optical
  Physics}\ }\textbf {\bibinfo {volume} {29}},\ \bibinfo {pages} {1489}
  (\bibinfo {year} {1996})}\BibitemShut {NoStop}%
\bibitem [{\citenamefont {Eland}(1987)}]{Eland1987}%
  \BibitemOpen
  \bibfield  {author} {\bibinfo {author} {\bibfnamefont {J.}~\bibnamefont
  {Eland}},\ }\href {\doibase 10.1080/00268978700101421} {\bibfield  {journal}
  {\bibinfo  {journal} {Molecular Physics}\ }\textbf {\bibinfo {volume} {61}},\
  \bibinfo {pages} {725} (\bibinfo {year} {1987})}\BibitemShut {NoStop}%
\bibitem [{\citenamefont {Hsieh}\ and\ \citenamefont
  {Eland}(1997)}]{ShizukaHsieh1997}%
  \BibitemOpen
  \bibfield  {author} {\bibinfo {author} {\bibfnamefont {S.}~\bibnamefont
  {Hsieh}}\ and\ \bibinfo {author} {\bibfnamefont {J.~H.~D.}\ \bibnamefont
  {Eland}},\ }\href {\doibase 10.1088/0953-4075/30/20/015} {\bibfield
  {journal} {\bibinfo  {journal} {Journal of Physics B: Atomic, Molecular and
  Optical Physics}\ }\textbf {\bibinfo {volume} {30}},\ \bibinfo {pages} {4515}
  (\bibinfo {year} {1997})}\BibitemShut {NoStop}%
\bibitem [{\citenamefont {Young}\ \emph {et~al.}(2018)\citenamefont {Young},
  \citenamefont {Ueda}, \citenamefont {G\"uhr}, \citenamefont {Bucksbaum},
  \citenamefont {Simon}, \citenamefont {Mukamel}, \citenamefont {Rohringer},
  \citenamefont {Prince}, \citenamefont {Masciovecchio}, \citenamefont {Meyer},
  \citenamefont {Rudenko}, \citenamefont {Rolles},\ and\ \citenamefont
  {et~al.}}]{AMO_roadmap}%
  \BibitemOpen
  \bibfield  {author} {\bibinfo {author} {\bibfnamefont {L.}~\bibnamefont
  {Young}}, \bibinfo {author} {\bibfnamefont {K.}~\bibnamefont {Ueda}},
  \bibinfo {author} {\bibfnamefont {M.}~\bibnamefont {G\"uhr}}, \bibinfo
  {author} {\bibfnamefont {P.~H.}\ \bibnamefont {Bucksbaum}}, \bibinfo {author}
  {\bibfnamefont {M.}~\bibnamefont {Simon}}, \bibinfo {author} {\bibfnamefont
  {S.}~\bibnamefont {Mukamel}}, \bibinfo {author} {\bibfnamefont
  {N.}~\bibnamefont {Rohringer}}, \bibinfo {author} {\bibfnamefont {K.~C.}\
  \bibnamefont {Prince}}, \bibinfo {author} {\bibfnamefont {C.}~\bibnamefont
  {Masciovecchio}}, \bibinfo {author} {\bibfnamefont {M.}~\bibnamefont
  {Meyer}}, \bibinfo {author} {\bibfnamefont {A.}~\bibnamefont {Rudenko}},
  \bibinfo {author} {\bibfnamefont {D.}~\bibnamefont {Rolles}}, \ and\ \bibinfo
  {author} {\bibnamefont {et~al.}},\ }\href {\doibase 10.1088/1361-6455/aa9735}
  {\bibfield  {journal} {\bibinfo  {journal} {Journal of Physics B: Atomic,
  Molecular and Optical Physics}\ }\textbf {\bibinfo {volume} {51}},\ \bibinfo
  {pages} {032003} (\bibinfo {year} {2018})}\BibitemShut {NoStop}%
\end{thebibliography}%

\end{document}